# Using the SOCIO Chatbot for UML Modelling: A Family of Experiments

Ranci Ren, John W. Castro, Adrián Santos, Oscar Dieste and Silvia T. Acuña

**Abstract**— **Context:** Recent developments in natural language processing have facilitated the adoption of chatbots in typically collaborative software engineering tasks (such as diagram modelling). Families of experiments can assess the performance of tools and processes and, at the same time, alleviate some of the typical shortcomings of individual experiments (e.g., inaccurate and potentially biased results due to a small number of participants). **Objective:** Compare the usability of a chatbot for collaborative modelling (i.e., SOCIO) and an online web tool (i.e., Creately). **Method:** We conducted a family of three experiments to evaluate the usability of SOCIO against the Creately online collaborative tool in academic settings. **Results:** The student participants were faster at building class diagrams using the chatbot than with the online collaborative tool and more satisfied with SOCIO. Besides, the class diagrams built using the chatbot tended to be more concise —albeit slightly less complete. **Conclusion:** Chatbots appear to be helpful for building class diagrams. In fact, our study has helped us to shed light on the future direction for experimentation in this field and lays the groundwork for researching the applicability of chatbots in diagramming.

---

# 1 INTRODUCTION

MODELLING is a fundamental part of the software development process, and it is often a collaborative activity, requiring the participation of stakeholders with different backgrounds and technical expertise [1]. Both asynchronous methods (i.e., based on version control) and synchronous mechanisms (e.g., based on online tools) are typically used for modelling purposes. Synchronous mechanisms include a recent plethora of cloud-based platforms, supporting real-time collaboration (e.g., Lucidchart, Gliffy, and Creately). The use of social networks is pervasive in this day and age, and they have been recognized as a tool with substantial potential for software engineering (SE) [2], [3]. The SOCIO chatbot is a collaborative modelling tool, developed by our colleague de Lara and his research group for building models or meta-models [4]. The chatbot is accessible from Twitter or Telegram (nick @ModellingBot). Thanks to the SOCIO chatbot, designers and stakeholders can take advantage of social network collaborativeness and ubiquity to perform lightweight modelling tasks [4].

Chatbots have become increasingly common in our daily lives. Tech giants regard chatbots as a cost-effective solution since they can handle a tremendous number of simple customer requests, whereas customers receive a pleasant, convenient, and fast service. For example, Amazon has Alexa, Google has Google Assistant, Microsoft has Cortana, and Apple has Siri. Usability is a critical issue for chatbots, which now deal with all sorts of activities, including, as already mentioned, SE tasks like modelling. It is essential to assess a chatbot's usability, as poor interaction would impact user willingness to continue to use the service [5]. Usability is defined as a subset of quality in use. According to ISO/IEC 25010:2011 [6], usability consists of the characteristics of effectiveness, efficiency and satisfaction. ISO 9241-11:2018 also describes this term as the extent to which a system can be used by specified users to achieve specified goals with effectiveness, efficiency and satisfaction in a specified context of use [7]. Usability and quality in use requirements are defined by the International Organization for Standardization's ISO/IEC/IEEE 29148:2018 as they can include measurable effectiveness, efficiency, satisfaction criteria and avoidance of possible harm [8].

Laboratory experiments are commonplace in both SE research and SE practice [9], [10], [11]. Experiments can assess the effectiveness of SE treatments (e.g., tools) and check whether hypotheses on the effectiveness of such treatments hold. Unfortunately, the results of isolated experiments may be unreliable due to the small number of subjects typically participating in SE experiments [12]. With the aim of increasing the reliability of the results of individual experiments, SE researchers are collaborating to build families of experiments by means of replication [9].

Families of experiments can surpass the sample size-related limitations of individual experiments and also evaluate the effects of the treatments in different settings [13]. Families provide certain advantages for evaluating the effectiveness of SE treatments [14], [15], [16], [17], [18]: (i) researchers can apply consistent pre-processing and analysis techniques (as researchers have access to the raw data) to analyse the experiments and, in turn, increase the reliability of joint results; (ii) researchers conducting families of experiments may opt to reduce the number of changes made across the experiments with the aim of increasing the

• R. Ren is with the Universidad Autónoma de Madrid, Madrid, Spain. E-mail: ranci.ren@uam.es
• J. W. Castro is with the Universidad de Atacama, Copiapó, Chile. E-mail: john.castro@uda.cl
• A. Santos is with the University of Oulu, Oulu, Finland. E-mail: adrian.santos.parrilla@oulu.fi
• O. Dieste is with the Universidad Politécnica de Madrid, Madrid, Spain. E-mail: odieste@fi.upm.es
• S. T. Acuña is with the Universidad Autónoma de Madrid, Madrid, Spain. E-mail: silvia.acunna@uam.es

(Corresponding author: John W. Castro).

internal validity of joint results; and (iii) joint results are not affected by the detrimental effects of publication bias because families do not rely on already published results.

In this paper, we report a family of experiments conducted in academic settings that build upon a baseline experiment already published at a conference [19]. This paper extends our previous research [19] by replicating the experiment for two new scenarios, thus building a family of three experiments. This extension has an impact on the contents of the research, as further experimentation, calculations and analysis have to be conducted. The family of experiments that we ran compared the usability of the SOCIO chatbot with Creately in order to increase the reliability of the results of the baseline experiment [19].

The aim of our family of experiments is to answer the following research question:

**RQ:** *Compared to Creately, does the use of SOCIO positively affect the efficiency, effectiveness, and satisfaction of the users building class diagrams and the quality of the class diagrams built?*

In response to this research question, we ran a baseline experiment [19], and two identical replications. We chose to conduct identical rather than differential replications (i.e., replications with different experimental set-ups or student participant characteristics) to output an accurate picture of SOCIO performance. We analysed the data with descriptive statistics, violin plots and profile plots. Finally, we combined and meta-analysed the data using linear mixed models [20]. The main contributions of this article are: (1) a family of experiments that provides evidence of the usability of the SOCIO chatbot, and (2) a list of suggestions from SOCIO chatbot users (collected in four open-ended questions using a satisfaction questionnaire) that may help us to understand the impact of three human-computer interaction (HCI) usability characteristics (effectiveness/efficiency/satisfaction) on collaborative modelling and chatbot design.

The paper is structured as follows. In Section 2, we provide a brief overview of the SOCIO and Creately tools. In Section 3, we discuss related work on usability experiments for chatbots. In Section 4, we describe the design of the family of experiments and the respective objectives, hypotheses, response variables and tasks, as well as the profile of the subjects. In Section 5, we report the data analysis and results of the family of experiments. Section 6 outlines validity threats. The paper ends with a discussion of the results (Section 7) and conclusions and future work (Section 8).

## 2 BACKGROUND

The SOCIO chatbot is a collaborative tool for creating class diagrams that is integrated into Twitter and Telegram (with nick @ModellingBot), thus benefitting from social media ubiquity and collaborativeness [4], [21]. This method enables users to undertake collaborative design anytime and anywhere, thereby increasing class diagram construction flexibility. Through the social network, messages can be addressed to SOCIO and interpreted and used to build the diagram. For example, the messages sent to

SOCIO via Telegram can be a descriptive command or message. Commands are imperative operations, like "add class X" or "set attribute size to int", that manipulate the diagram directly. Descriptive messages are natural language (NL) sentences on the context of the diagram to be built, such as "the shop contains products". To mitigate possible NL processor errors highlighted by previous research [21], SOCIO provides for models to be manipulated directly using NL (i.e., employing sentences like "remove employee", which are parsed and interpreted) and actions to be undone. Additionally, messages in micro-blogging systems are short, which facilitates NL processing. As shown in Fig. 1, two participants (i.e., Anonymous A and Anonymous B) form a pair of designers, the chatbot SOCIO responds to their natural language commands with the latest change in the class diagram (highlighted in green within a green rectangle). Fig. 1 illustrates a communication scenario with the SOCIO chatbot and collaboration between the pair of designers.

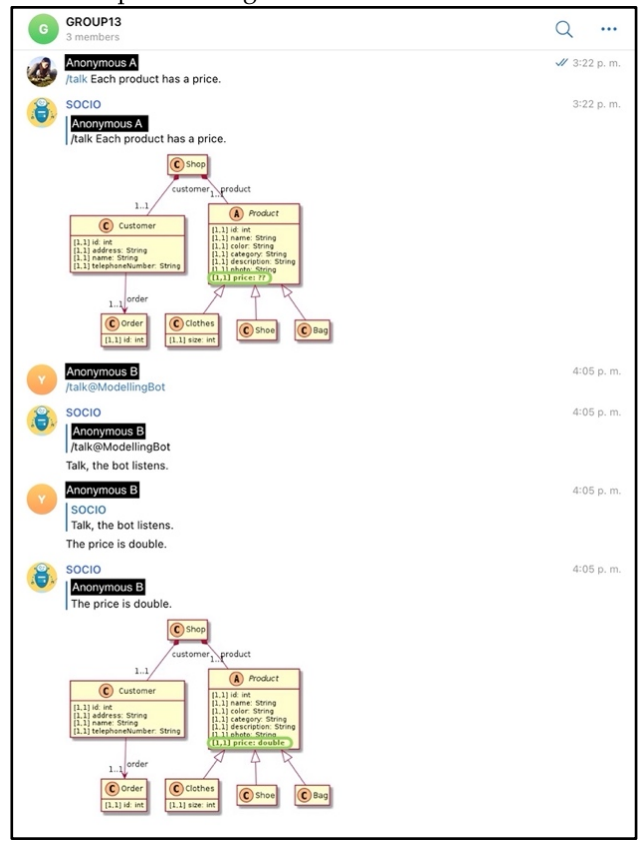

Fig. 1. SOCIO collaboration example.

Creately (https://creately.com/app) is a real-time collaborative tool, with a friendly interface and learnability, built on Adobe's Flex/Flash technologies that can create over 50 types of diagrams, including class diagrams. It provides the drag-and-drop feature and the diagram components required to build class diagrams. As shown in Fig. 2, three participants (i.e., Anonymous A, Anonymous B, Anonymous C) form a team.

Fig. 2 shows a team collaboration scenario provided by one of participated teams via Creately (i.e., this diagram may not be completely correct).

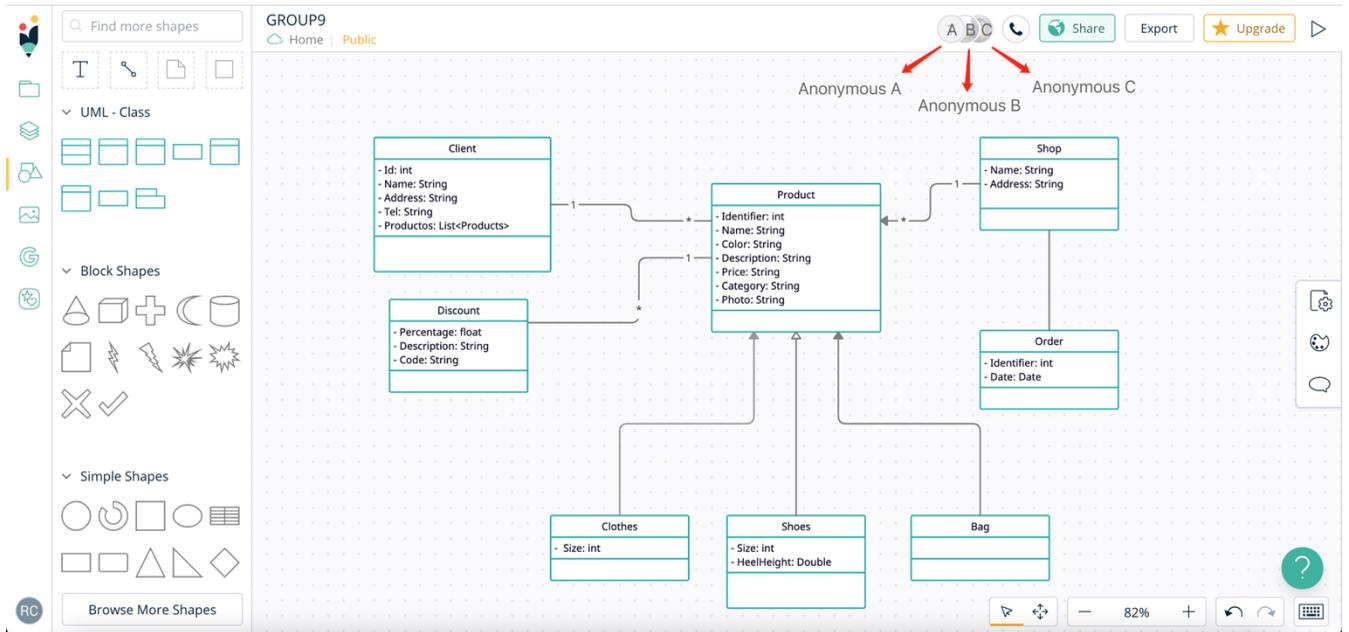

Fig. 2. Creately collaboration example.

Additionally, the tool supports offline work and resynchronization when the internet connection is available. We chose Creately as the control tool since there were no previous studies assessing the usability of Creately, and it is the most used online collaborative modelling tool [22] with similar functionality to SOCIO.

In contrast to SOCIO, however, Creately uses a traditional graphical user interface (GUI), accessible through a web browser. While SOCIO embeds modelling into social networks, Creately lacks embedded chat. Therefore, it has to resort to external options such as Telegram.

## 3  RELATED WORK

Ren et al. [23] reported a wider systematic mapping study (SMS) to identify the state of the art with respect to chatbot usability and applied HCI techniques in order to analyse how to evaluate chatbot usability. The SMS search ranged from January 2014 to October 2018. They concluded that chatbot usability is a very incipient field of research, where the published studies are mainly surveys, usability tests, and rather informal experimental studies. Hence, it is necessary to perform more formal experiments to measure user experience and exploit these results to provide usability-aware design guidelines. We updated the SMS, focusing on papers published from November 2018 to June 2021. Additionally, we looked at papers published at HCI conferences and in HCI journals from 2014 to 2021 to assure that we did not to miss important papers. We found 10 related papers from HCI conferences and five related papers from HCI journals. After applying inclusion and exclusion criteria for the retrieved papers, three articles from HCI journals and two papers from HCI conferences were initially added into our candidate papers. However, three journal articles were duplicates of papers identified by our previous SMS. Finally, two HCI conference papers from these resources were added to our primary studies as [24] and [25]. To give readers a better understanding of the SMS, we report the procedure in the supplementary material and describe the validity threats to the SMS in Section 6.

Our SMS basically followed the SMS process reported by Ren et al. [23]. Since this is an extension of previous research, we used the same search string as [23] (("usability" OR "usability technique" OR "usability practice" OR "user interaction" OR "user experience") AND ("chatbots" OR "chatbots development" OR "conversational agents" OR "chatterbot" OR "artificial conversational entity" OR "mobile chatbots")). Unlike the above SMS [23], we review chatbot usability evaluation experiments in order to identify recent trends and methodologies in the experimental software engineering field. The selection criteria used to retrieve the primary studies are summarized below.

Inclusion criteria:
- The abstract or title mentions an issue regarding chatbots and usability; OR
- The abstract mentions an issue related to usability engineering or HCI techniques; OR
- The abstract mentions an issue related to user experience; AND
- The paper describes a controlled experiment on chatbot usability.

Exclusion criteria:
- The paper presents only an evaluation or a quasi-experiment related to chatbot usability; OR
- The paper does not present any controlled experiment related to chatbot usability; OR
- The paper does not present any issue related to the chatbots and usability; OR
- The paper does not present any issue related to the chatbots and user interaction; OR
- The paper does not present any issue related to chatbots and user experience; OR
- The paper is written in a language other than English.

Finally, we retrieved 25 primary studies related to experiments on the usability of chatbots analysed in this study. These experiments measured usability characteristics referred to effectiveness, efficiency and satisfaction as shown in Table 1. Some of the primary studies ([19], [26], [27], [28], [29], [30], [31], [32], [33]) consider all three of these aspects, others ([24], [25], [34], [35], [36], [37], [38], [39], [40]) investigate efficiency and satisfaction, others again ([41], [42], [43], [44], [45]) consider only satisfaction, whereas only one ([46]) evaluated both effectiveness and satisfaction.

TABLE 1
USABILITY CHARACTERISTICS IN PRIMARY STUDIES

| Ref. | Characteristics of Usability |
| --- | --- |
| [19] | **Effectiveness:** Task completion<br>**Efficiency:** Task completion time; Communication effort<br>**Satisfaction:** User experience |
| [24] | **Efficiency:** Task completion time<br>**Satisfaction:** Ease of use; User experience; Helpfulness; Attentiveness |
| [25] | **Efficiency:** Task completion time; Communication effort; Response quality<br>**Satisfaction:** User experience; During use; Valuable |
| [26] | **Effectiveness:** Task completion<br>**Efficiency:** Task completion time<br>**Satisfaction:** Namely pragmatic quality; Hedonic quality; Attractiveness |
| [27] | **Effectiveness:** Accuracy; Precision<br>**Efficiency:** Task completion time; Mental effort; Communication effort<br>**Satisfaction:** Ease of use; Complexity control; Pleasure; Want to use again; Learnability; Adaptability |
| [28] | **Effectiveness:** Number of errors/error rate<br>**Efficiency:** Task completion time; Mental effort<br>**Satisfaction:** Ease of use; Complexity control |
| [29] | **Effectiveness:** Number of errors/error rate<br>**Efficiency:** Response quality<br>**Satisfaction:** Learnability; Semantic intelligence; Clarity/details; Recognition over recall; Mapping |
| [30] | **Effectiveness:** Number of errors/error rate<br>**Efficiency:** Response quality<br>**Satisfaction:** Ease of use; Pleasure; Learnability; User experience |
| [31] | **Effectiveness:** Accuracy; Precision<br>**Efficiency:** Communication effort<br>**Satisfaction:** Ease of use; Want to use again/intent to use; User experience |
| [32] | **Effectiveness:** Task completion<br>**Efficiency:** Task completion time<br>**Satisfaction:** Overall user experience |
| [33] | **Effectiveness:** Task completion<br>**Efficiency:** Task completion time; Mental effort<br>**Satisfaction:** Ease of use; Pleasure; Want to use again |

TABLE 1
USABILITY CHARACTERISTICS IN PRIMARY STUDIES (CONT'D)

| Ref. | Characteristics of Usability |
| --- | --- |
| [34] | **Efficiency:** Task completion time; Communication effort<br>**Satisfaction:** Ease of use; Context-dependent question; Complexity control; Pleasure; Want to use again; Learnability |
| [35] | **Efficiency:** Task completion time<br>**Satisfaction:** Ease of use; Pleasure; User experience |
| [36] | **Efficiency:** Task completion time<br>**Satisfaction:** Ease of use; Before use; During use; Pleasure; User experience |
| [37] | **Efficiency:** Response quality<br>**Satisfaction:** Ease of use; Pleasure; Want to use again; Learnability |
| [38] | **Efficiency:** Task completion time; Communication effort<br>**Satisfaction:** Pleasure; Learnability; User experience; Pragmatic quality; Valuable |
| [39] | **Efficiency:** Task completion time; Response quality<br>**Satisfaction:** User experience; Attractiveness; Want to use again/intent to use; Chatbot recommendation to third parties |
| [40] | **Efficiency:** Mental effort<br>**Satisfaction:** During use |
| [41] | **Satisfaction:** Pleasure |
| [42] | **Satisfaction:** Ease of use; Valuable; Chatbot recommendation to third parties |
| [43] | **Satisfaction:** User experience |
| [44] | **Satisfaction:** Ease of use; Pleasure; Learnability; Overall user experience |
| [45] | **Satisfaction:** Ease of use; Pleasure; Want to use again; Learnability |
| [46] | **Effectiveness:** Experts and users' assessment<br>**Satisfaction:** Learnability |
| [47] | **Effectiveness:** Number of errors/error rate<br>**Satisfaction:** User experience |

Satisfaction is again the usability characteristic of most concern to researchers since it is evaluated most often. We noticed that chatbot designers tend to evaluate chatbot usability with a view to putting them into use in real life or in the industrial field [34]. The chatbot in [27] is equipped with a sentiment analyser, as it discovers items that best fit user needs, whereas the conversational agent developed in [41] dynamically adjusts its conversation style, tone and volume in response to users' emotional, environmental, social and activity context.

With regard to efficiency, we find that there are studies focusing on measuring task completion time, for example, the time spent on each individual question and the number of concepts that the user can introduce for each chatbot message [27]. The hedonic quality of conversation is relevant to chatbot efficiency, since the effort required of users to understand and answer a chatbot request is frequently measured. In conclusion, researchers have, in recent years, been trying to understand chatbot reaction time and clarity of speech.

As far as effectiveness is concerned, we observed that task completion and number of errors/error rate are the characteristics of most concern. The researchers measured the accuracy and precision of the chatbot response [27] and the number of errors that occurred in communication between the subject and the chatbot [28].

Within the primary studies, chatbots are dedicated to various domains. Most chatbots are used as personal assistants [25], [26], [29], [30], [34], [37], [38], [39], [40], [42], [43], [44], [45], [46], [47], especially in the healthcare domain [32], [35], [38], [44], some act as recommenders [27], [28], [31], emotion-aware conversational agents [41] and e-commerce chatbots [33], [36]. Nevertheless, none of the above chatbots is applied as a modelling tool like the SOCIO chatbot.

Appendix A outlines the 25 experiments on chatbot usability, including information like the goal, article research questions (ARQ), number of experiments in the family, sample sizes and type of subjects. Of the usability experiments that we reviewed, only one study conducted replications of experiments. This study was designed as a within-subjects mixed-method experiment with participants from different regions or backgrounds [42]. In view of this, we executed a family of experiments on chatbot usability with the hope of filling research gaps.

Note that one study [43] conducted two different experiments in controlled laboratory-based and real-world environments to comprehensively evaluate the usability of their chatbot. Since the experimental designs are different, we do not consider this study to be a family of experiments.

Of the 18 experiments that defined the type of subject, seven recruited students as subjects, i.e., they used an academic setting for their experiments. Regarding the research questions and research goals, 68% (17) of the experiments compared the chatbot with another tool with similar functions, while 28% (7) of the experiments compared different functionalities or approaches of the same chatbot. Only one experiment performed an analysis of scaled questions and semantic analysis, and no comparisons were made.

Here we provide an overview of the research that has been conducted so far on the usability of the SOCIO chatbot. They include the baseline experiment of the family of experiments reported here [19], and two individual evaluations [4], [21]. All the participants in all three experiments are students with a computer science background. They were asked to fill in the questionnaire after completing the experimental task in each case. The experiments reported in [4], [21] are two small-scale SOCIO evaluation experiments (with 19 and 8 participants, respectively). In [4], the researchers measured the suitability of this chatbot. In [21], a consensus mechanism for choosing different modelling alternatives was assessed. The tasks in both experiments were carried out in groups. However, these two experiments focused on evaluating SOCIO separately (i.e., with no comparison to an alternative treatment), and the tasks were simple. In the first evaluation [4], the researchers recruited four groups of participants to assess the suitability of the SOCIO chatbot by building a class diagram for electronic commerce in a maximum of 15 minutes. They evaluated SOCIO on four aspects: (1) suitability of NL for modelling with respect to the use of an editor, (2) chatbot NL interpretation precision, (3) command set functionality sufficiency, and (4) whether users liked having a modelling tool embedded in a social network or would prefer a separate collaborative tool. The result suggests that participants liked the idea of collaborating through social networks, and natural language was found to be useful for task completion. In the second assessment [21], the researchers focused on evaluating the SOCIO chatbot group consensus mechanism. During the experiment, participants were required to choose, after a short tutorial, the best of three options for two projects, with or without the consensus mechanism. The five-point Likert scale results show that the consensus mechanism was considered especially useful for large groups (average 4.7/5).

Our baseline experiment [19] compared the SOCIO chatbot with Creately, a traditional web-based application, evaluated by a larger number of users (54 student participants divided into 18 three-member teams). This experiment adopted a cross-over design [48].

In this study, we conducted two identical replications of the above baseline experiment to build a family of experiments in academic settings. We aggregated the results of this family of experiments to output more reliable results than yielded by the baseline experiment alone thanks to a larger sample size.

## 4 FAMILY DESIGN

This section describes the design of our family of experiments.

### 4.1 Objectives, Hypotheses and Variables

We state the aim of our family of experiments as follows: evaluate, by means of controlled experiments, the usability of the SOCIO chatbot compared with the Creately web tool in terms of user effectiveness, efficiency and satisfaction and the quality of the resulting class diagrams in academic settings. Note that our aim is to identify the aspects that can improve chatbot usability, and particularly SOCIO chatbot usability, rather than building better class diagrams either individually or collaboratively in small teams.

The null hypotheses governing this research question are:

H.1.0 There is no significant difference in efficiency using SOCIO or Creately when building the class diagram.

H.2.0 There is no significant difference in effectiveness using SOCIO or Creately when building the class diagram.

H.3.0 There is no significant difference in satisfaction using SOCIO or Creately when building the class diagram.

H.4.0 There is no significant difference in the quality of the class diagram built using SOCIO or Creately.

The main independent variable across all experiments is the modelling tool, with either the SOCIO chatbot or the Creately online application (i.e., two tools with two similar functions) as treatments. The response variables within the family are usability characteristics and class diagram quality. We measure usability as efficiency, effectiveness and satisfaction. Based on definitions from ISO/IEC 25010:2011 [6], ISO 9241-11 [7], ISO/IEC/IEEE 29148 [8] and Hornbæk's guide [49], efficiency, effectiveness and satisfaction are commonly measured characteristics for evaluating product usability. We also measure the quality of the class diagrams generated by the teams as a measure of effectiveness, in particular "quality of outcome" according to [49]. Note that, although "quality of outcome" is a sub-characteristic of effectiveness, it is addressed separately for clarity. Specifically, we measured efficiency as follows:

- *Time:* Time, measured in minutes, taken by a team to complete the task (with a maximum of 30 minutes). To establish the time limit, we did a rehearsal with a team composed of three SE experts and found that it takes 10 minutes to complete each task. We assumed that each team in the real experiments would be composed of three novices, and we set the time to 10*3 minutes. On the other hand, we set the time limit to 30 minutes to prevent fatigue in the two-session experiment. During the experiments, experimenters enforced the time limit orally. Due to geographical and financial constraints, some of our experiments (EXP1 and five teams in EXP3) were conducted in Ecuador remotely via video, whereas the experimenters who oversaw the experiment were based in Spain. In Ecuador, however, a professor helped to enforce the time limit. The other experiments (EXP 2 and 11 teams in EXP3) were conducted in person, and the experimenters personally enforced the time limit. Regarding task completion time, we recorded the start (and stop) time of the Telegram chat, then we calculated the task completion time manually.

- *Fluency:* Number of discussion messages generated by a team during task completion via a Telegram group. Note that when we counted discussion messages, we only calculated the messages related to task performance, tool use and topics related to team management. It would be useless to count messages that did not refer to tools. We did not count messages stating that users were unhappy with a tool, jokes or questions put to experimenters. Notice that, although SOCIO and Creately have similar functionalities and the interaction takes place via Telegram in both cases, the messages have different characteristics. In addition to user discussion messages, SOCIO receives and processes commands, and generates error and success messages. On the other hand, Creately is a generic web-based diagramming tool, and Telegram is just external software that supports real-time communication between users. Therefore, Creately does not accept command messages or generate responses. Messaging in Creately is confined to discussion messages. Therefore, SOCIO and Creately have only one message type in common: discussion messages. These messages are represented by the discussion messages response variable.

We measured effectiveness as completeness, based on the comparison of each class diagram with the ideal class diagram (see Fig. 3 and Fig. 4) that we (i.e., the experimenters) built to measure the solutions produced by all participants [19], [49]. See the supplementary material for details. The participants had no access to these test oracles. The metrics of time, fluency, and completeness refer to social complexity and sociability and are typically evaluated when measuring macro-level usability (tasks requiring hours of collaboration) [6], [7], [49].

To assess and quantify satisfaction, we modified the System Usability Scale (SUS) questionnaire [21], [50] to suit our experiments. The modified SUS questionnaire consists of 10 SUS questions measured on a five-point Likert scale ranging from strongly disagree to strongly agree and three or four open-ended questions: we did not ask participants which tool they preferred until the end of the second period of each experiment. Finally, we adopted Brooke's equation [50] to derive the numerical value of each participants' satisfaction score as shown below. We selected the median of the scores given by the three members of each team for each questionnaire item as the team score:

$$SUS\ score = \left[\sum_{n=1}^{5}(P_{2n-1}-1)+(5-P_{2n})\right] \times 2.5, \quad (1)$$

where $P_{2n-1}$ are odd-numbered questions and $P_{2n}$ are even-numbered questions.

We took an ideal class diagram as a reference to measure the quality of each team's class diagram. However, there is more than one possible UML class diagram for any one modelling task, all of which can be considered as equally "correct". The ideal class diagram was designed by software engineering experts before the experiment took place. We used the following metrics to measure quality [51]:

**Precision** = TP / (TP + FP)           (2)
**Recall** = TP / (TP + FN)           (3)
**Accuracy** = (TP + TN) / (TP + TN + FP + FN)   (4)
**Error** = (FP + FN) / (TP+ TN + FP + FN)     (5)
**Success** = TP / (#Number of ideal class diagram elements)           (6)

The above formulas can be computed by comparing each team's class diagram with the ideal class diagram:

- TP (true positive): Number of elements that are found in both the ideal class diagram and the team class diagram.

- FN (false negative): Number of elements that are found in the ideal class diagram, but not in the team class diagram.

- FP (false positive): Number of elements that are found in the team class diagram, but not in the ideal class diagram.

- TN (true negative): There are no true negatives as a result of model comparison, and hence the value is always 0.

Precision provides the percentage of correct classes in each team's solution based on the elements of the ideal diagram. Recall is a completeness metric, giving the percentage of classes of the ideal diagram contained in the solution.

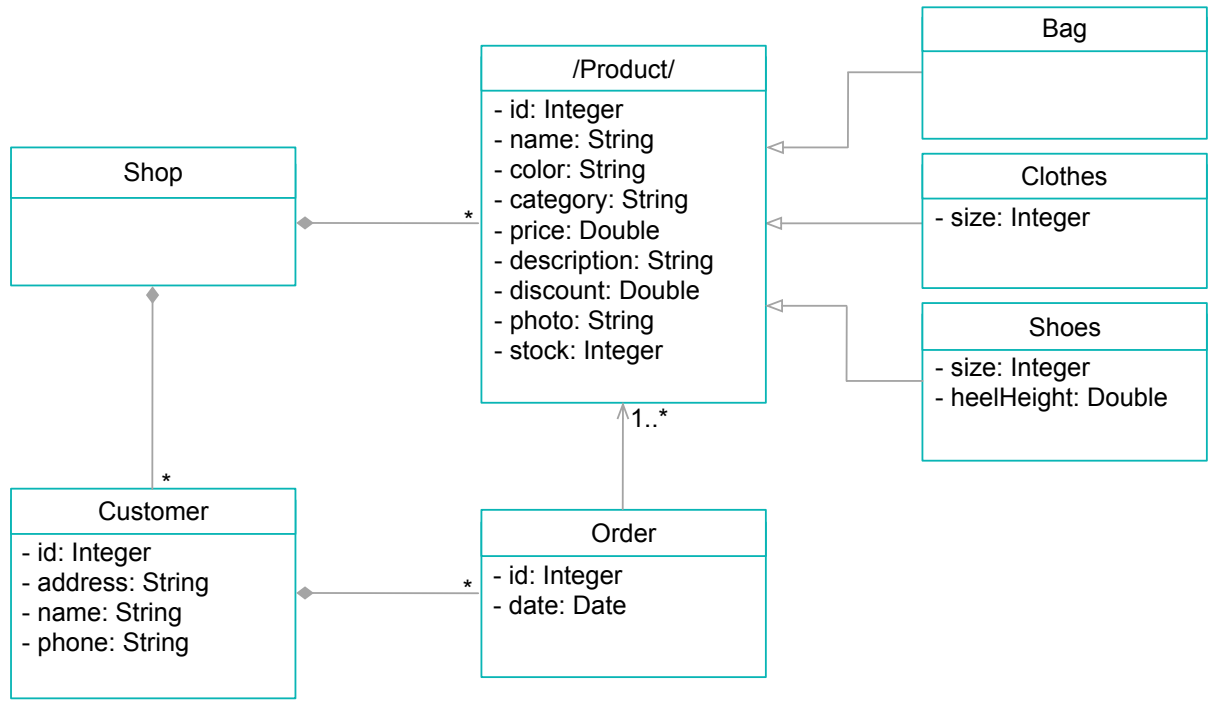

Fig. 3. Ideal solution for Task 1.

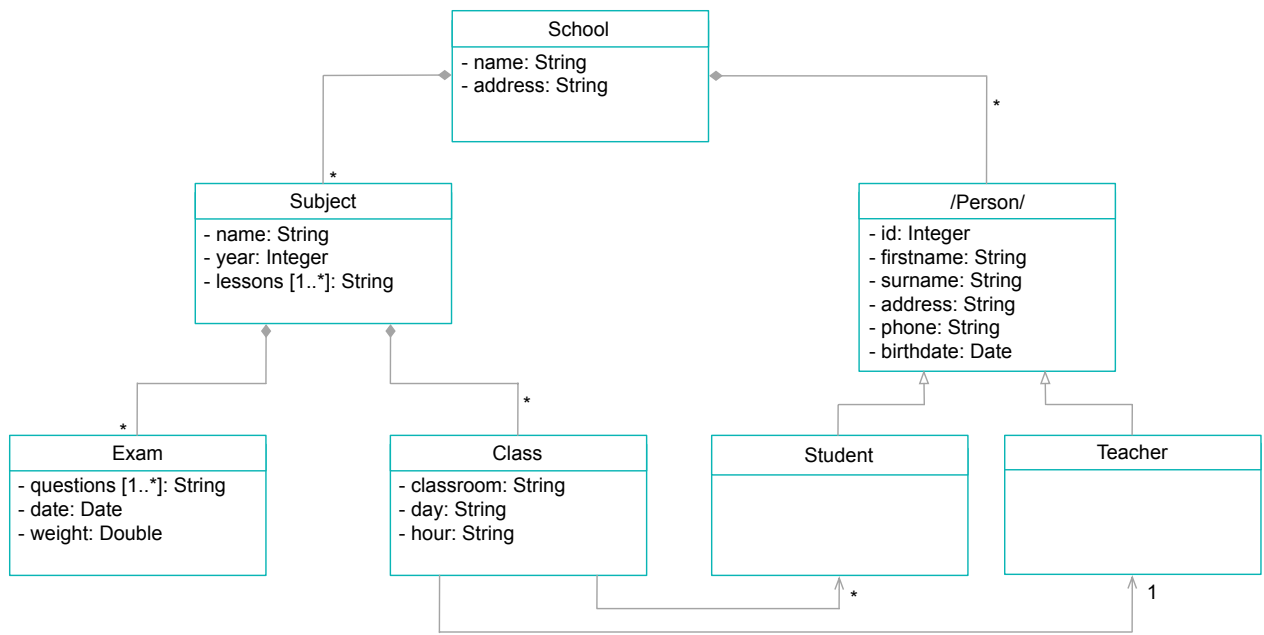

Fig. 4. Ideal solution for Task 2.

Accuracy combines both of the above metrics: it is the ratio of correctly developed elements to all elements found and not found in the ideal class diagram. Error reflects unreliability, that is, how many elements are redundant or missing in each solution. Perceived success refers to the success rate of each team, directly compared with the ideal class diagram. Their respective formulas have been computed by comparing the ideal class diagram with each class diagram built. They are reported in [19].

## 4.2 Design of the Experiments

We ran a total of three identical replications, which we grouped to form this family of experiments in academic settings. We decided to conduct identical rather than differential replications because of the relatively small sample size of our baseline experiment (i.e., 18 subjects [19]) and the resulting potential for inaccurate and/or biased results [52].

As SOCIO is a new tool, we thought it was sensible to first increase the sample size of the original experiment (in order to output more accurate results) and then start making changes across the replications in a later iteration to study the impact of such changes on results [13]. We decided to increase the sample size of the original experiment by replicating it as closely as possible with different groups of students. The power analysis reported in Appendix B shows that we need 50 teams to achieve 80% power across all response variables. This having been said, only two response variables (discussion messages and satisfaction) need such large numbers. Most of the response variables are in the range of 10-35 teams. Altogether, the three experiments comprised 44 teams.

Our rationale for pursuing this approach of running identical replications is that if we had changed the design of the experiments with such small sample sizes, we would not have been able to distinguish whether the results were due to the changes made or to the relatively small sample sizes of the experiments (and the resulting large fluctuations to be expected in the results) [52]. This may have had distorted the results of the aggregation [52].

We designed all the experiments as two-sequence and two-period within-subjects cross-over designs (see Table 2). We chose a within-subjects design rather than a between-subjects design to: (1) increase the effective number of data points and thus improve the statistical power of the experiments; and (2) reduce the variability of the data across the treatments (as each subject acts as its own baseline in within-subjects designs). Instead of a simple within-subjects design, we opted for a cross-over design to: (1) avoid the influence of the period on treatment; and (2) avoid learning effects from one treatment to another [48].

TABLE 2
EXPERIMENTAL DESIGN

| Group | Period 1 (Task 1) | Period 2 (Task 2) |
| --- | --- | --- |
| Group 1 (SC-CR) | SOCIO | Creately |
| Group 2 (CR-SC) | Creately | SOCIO |

Regarding team size, Dyer defines that a team consists of at least three people, each of whom should perform a specific assigned role or function and where the achievement of the goal requires some form of dependency among group members [53]. Additionally, the number of individuals in a team affects performance in many ways, for instance, when the team is very large, the individual efforts of members are more likely to be overshadowed by the efforts of the team [54]. Team size also depends on the needs of the task [55]: while more team members provide more resources to complete a task, smaller teams are more efficient [56]. Since the tasks required in our experiment are not very complicated, we chose a team size of three members in order to maximize the subject size and get the highest possible statistical power. Besides, we focus on teamwork because design discussions today are usually conducted in a collaborative teamwork environment.

Therefore, the participants were assigned to three-member teams. Each team created a class diagram using the SOCIO chatbot (SC) and the Creately web application (CR). We randomly assigned each team to one out of two groups

(Group 1 or Group 2). Accordingly, each group applies the treatments in a different order (SC-CR / CR-SC) on a particular task.

After a 10-minute introduction about the tool that the participants were to use before each period, they were asked to perform the task with the tool in a maximum of 30 minutes. Group 1 implemented Task 1 with SOCIO in the first period and then Task 2 with Creately in the second period (i.e., SOCIO-Creately sequence). Conversely, Group 2 implemented Task 1 with Creately in the first period and then Task 2 with SOCIO in the second period (i.e., Creately-SOCIO sequence). Upon completion of each experimental task, all participants filled in a SUS questionnaire associated with the tool that they had just used (i.e., all participants had to fill in the modified SUS questionnaire twice with respect to two modelling tools). The participants were also asked if they preferred SOCIO or Creately in the second SUS questionnaire.

We designed two different experimental tasks (each assigned to a different experimental period). The first task is to develop a class diagram for a store including the management of products and customers. The second task is to design a class diagram for a school to support courses and students. We adapted the complexity of the class diagrams to the length of the experimental periods. During the experiment, members of the same team were allowed to communicate with each other via Telegram groups only. This aimed to ensure that all the experimental data were recorded. In all the experiments, the participants had a maximum of 30 minutes to develop each task in order to prevent fatigue.

## 4.3 Subjects

The participants in our family of experiments were students recruited at two universities in two countries: (1) the *Universidad de las Fuerzas Armadas ESPE Extensión Latacunga* (ESPE-Latacunga) in Ecuador (UNIV-1), and (2) the *Escuela Politécnica Superior of the Universidad Autónoma de Madrid* (EPS-UAM) in Spain (UNIV-2).

We grouped the participants in three-member teams by means of a random team generator (www.random-lists.com/team-generator). All the participants (i.e., a total of 132) were grouped in a total of 44 different teams (i.e., subjects). Each team is considered to be one subject.

Table 3 shows the overview of subjects. EXP1, the baseline experiment, was run with 18 subjects (54 participants) from UNIV-1, EXP2 was run with 10 subjects (30 participants) from UNIV-2, and EXP3 was run with 11 subjects from UNIV-2 and 5 subjects from UNIV-1 (48 participants in total).

TABLE 3
OVERVIEW OF SUBJECTS

| EXP | Time | Affiliation | #Participants | Teams |
| --- | --- | --- | --- | --- |
| EXP1 | Jul, 2019 | UNIV-1 | 54 | 18 |
| EXP2 | Jul, 2019 | UNIV-2 | 30 | 10 |
| EXP3 | Nov, 2019 | UNIV-1 | 15 | 5 |
|  |  | UNIV-2 | 33 | 11 |
| Total |  |  | 132 | 44 |

All the participants were undergraduate students who were completing a computer engineering degree. All the participants volunteered to participate in exchange for some stationery.

Participants did not receive any training or participate in a trial session before the experiment was run. At the very beginning of the experiments, all participants signed an informed consent form indicating that they granted us permission to record their data via Telegram. We ensured that the participation of all the subjects in this research was completely anonymous and that no information they shared could be electronically traced to them. Before participating in the experiment, the subjects completed a questionnaire that assessed some demographic data and their related experience and knowledge (i.e., level of English, preconceived ideas regarding their social media and chatbot use, and level of knowledge on class diagrams). In particular, we designed ordinal-scale questions to collect and measure the subjects' knowledge of English and Telegram (required) and chatbot and class diagramming familiarity, ranging from 1 (no knowledge or novice) to 5 (very familiar or expert) in order to rate their level of perceived awareness.

According to the familiarity questionnaire, we collected basic information about all the participants. They have the following characteristics:

- The final sample consists of 132 participants. Of the sample, 95 are men and 37 are women.
- Participants have a mean age of 22 and a standard deviation of 0.12. The highest concentration of participants is in the range of 22 to 23 years.
- 86% of participants use social media frequently. The social media most used by participants are WhatsApp, Facebook, Instagram and Telegram.
- 62.9% of the participants have used or frequently use the Telegram application, while 37.1% have never used this app.
- 83.3% of participants believe they are relatively familiar with class diagrams.
- 31.82% of participants have never used a chatbot, while 68.18% have related experience (55.3% have used chatbots from time to time and 12.88% are regular users).

Although the universities participated in more than one experiment, we ensured that each participant participated in our experiments only once. They were recruited from different classes and degree programmes. All 132 participants had taken the Software Design and Analysis Project course. Therefore, they had the knowledge of class diagrams that they needed to carry out the experimental tasks. Fig. 5 shows the profile plot of the participant characteristics across the replications: Telegram, social media and chatbot experience; knowledge of chatbots and class diagramming, and English level.

As Fig. 5 shows, there are observable patterns among averaged participant-level characteristics of replications: EXP1 participants claimed to be relatively more experienced in terms of Telegram and chatbot use and more knowledgeable about class diagrams and chatbots than

EXP2 and EXP3 subjects, while EXP3 participants have better English language skills. Participants' average experience looks to be heterogeneous, but the gaps between each experiment appear to be small (i.e., they are less than 1).

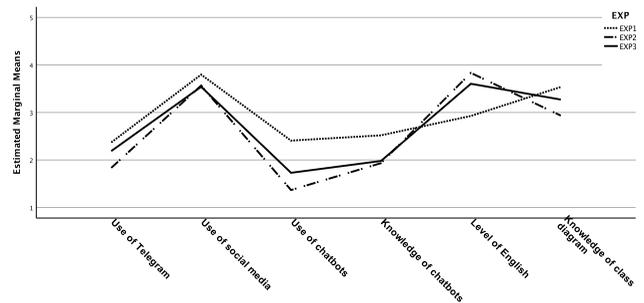

Fig. 5. Profile plot for subjects' experience.

Summing up the above analysis, the sample of the participants in our family is, according to the results of the familiarity questionnaire and our perceptions while running the experiments, knowledgeable about class diagrams. Participants claimed that they understood what chatbots are –at least at the conceptual level–, while they considered that they had more knowledge of than experience with chatbots.

With regard to the experimental platform, all participants are obliged to use and communicate using Telegram. Although 37.1% of the participants had no experience with this app, no complicated operations are required to perform the tasks, and subjects were in fact experienced in social media use to ensure that they would be able to successfully complete the experiment. Although none of the subjects are native English speakers, they all considered that they had at least an intermediate level of English. As there are no significant differences in age, gender, knowledge background, social media usage habits, smartphone or tablet ownership across the three experiments, we consider that the participants are comparable even though they come from different countries.

## 5 RESULT AND AGGREGATION OF DATA

In this section, we answer the research question: Compared to Creately, does the use of SOCIO positively affect the efficiency, effectiveness, and satisfaction of the users building class diagrams and the quality of the class diagrams built? Here we follow Santos et al.'s [13] guidelines to analyse the family of experiments.

### 5.1 Analysis Approach

In the following, we analyse one by one each of the response variables (i.e., efficiency, effectiveness, satisfaction and quality). In particular, we focus on their respective metrics (i.e., time and discussion messages for efficiency; completeness for effectiveness; satisfaction for satisfaction; and precision, recall, accuracy, error and perceived success for quality).

For each metric we provide: (i) descriptive statistics and violin plots divided by treatment (i.e., SOCIO, Creately) and by experiment (i.e., EXP1, EXP2 and EXP3), (ii) a profile plot showing the mean effect of the treatments across the experiments, and (iii) the joint results of all the experiments together by means of a one-stage individual participant data (IPD) meta-analysis and the contrast between treatments across the experiments [57].

The descriptive statistics and violin plots ease the understanding of the data in each experiment. The profile plots give a bird's eye view of the data at family level and check for the existence of patterns across the results [13]. We followed an IPD meta-analysis approach rather than a meta-analysis of effect sizes because we had access to the raw data of the experiments [57].

As all the experiments have an identical (i.e., a crossover) design, they are analysed as advised by Vegas et al. [48]. In particular, we analyse the experiments using linear mixed models (LMMs) [9]. We used LMMs rather than their non-parametric counterparts because: (1) commonly used non-parametric models cannot be used to study the effect on the outcomes of multiple factors (e.g., period, treatment, and sequence ) at the same time; (2) the overall sample size (i.e., 44 teams, each with two data-points —one per session, a total of 88 data points) may suffice to make the central limit theorem hold [58], and, thus, interpret the results despite slight departures from normality, and (3) LMMs have fewer requirements than other methods, e.g., compound symmetry in the case of repeated-measures ANOVA.

In particular, we fit a three-factor LMM [59] for each metric: period (i.e., 1 or 2), treatment (i.e., SOCIO or Creately), and sequence (i.e., SOCIO-Creately or Creately-SOCIO). We add an extra parameter to the LMM to account for the difference of results across the experiments (i.e., Experiment). This is a common feature of stratified individual participant data (IPD) models [13]. We interpret the statistical significance of the results with the corresponding ANOVA table of LMMs. We estimated statistical power with a 0.05 significance level. However, the total number of teams that participated in the family of experiments is on the boundary of the number required to achieve 80% statistical power with a 0.05 significance level (see Appendix B). On this ground, we also considered any findings with a 0.1 significance level. We specify the significance level used together with each finding.

We round out the results of the LMMs with a cumulative meta-analysis [60] of LMMs (see Appendix C), which shows that the treatment estimates for both SOCIO and Creately tend to come closer (with increasingly narrower confidence intervals) as the experiment results within the family pile up (i.e., EXP1 vs. EXP1+EXP2, EXP1+EXP2+EXP3).

## 5.2 Response Variables
In the following we analyse the response variables one by one.

### 5.2.1 Efficiency
We measured efficiency in terms of *time* and *fluency*. Time corresponds to the time taken to complete the tasks. Fluency corresponds to the number of discussion messages exchanged by teammates during task development.

Figs. 6 and 8 show the profile plot for the mean value of time, and fluency, respectively, across the experiments. Figures 7 and 9 show the violin plot for time and fluency in all the experiments. The respective summaries of descriptive statistics are shown in Tables 4 and 7, grouped by experiment and treatment.

*Time.* As Figs. 6 and 7 show, the aggregate task completion time with SOCIO appears to be smaller than for Creately in two out of three experiments. The descriptive statistics also highlight this finding. As Table 5 shows, the difference in performance between the treatments is statistically significant (p-value<0.05). According to the pairwise contrast between the treatments in Table 6, **the participants took an average of 1 minute and 8 seconds longer to complete the task using Creately than with SOCIO**.

The effect may not appear to be very relevant (1 over 25 minutes, that is, a 4% improvement on average). However, quite a lot of operations can be performed on the class diagram in one minute. The chat history shows that users had time to generate 6 to 10 messages in the SOCIO Telegram group or make five modifications or movements of a class diagram in Creately in one minute. Furthermore, this time should be considered with respect to the completion of what can be regarded as rather simple experimental tasks. It is to be expected that the effect would be greater for more complex tasks. Further experiments would be required to check this suspicion.

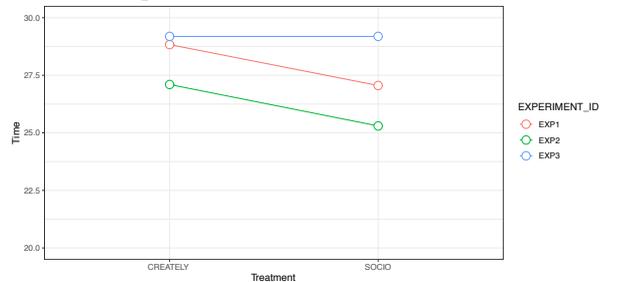

Fig. 6. Profile plot for time spent on tasks.

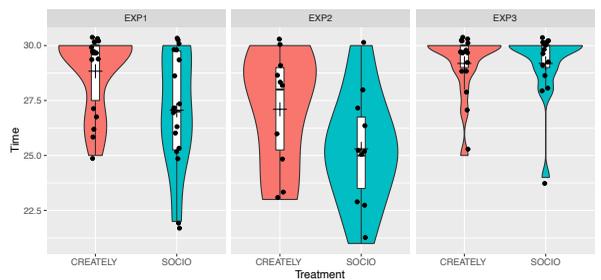

Fig. 7. Violin plot for time spent on tasks.

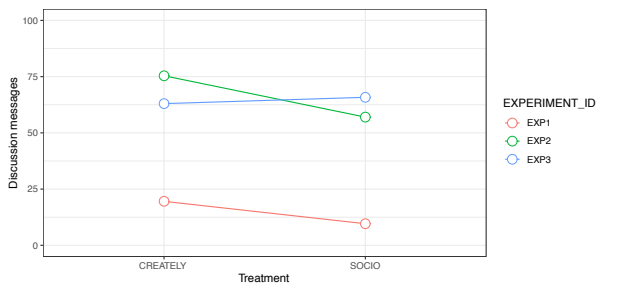

Fig. 8. Profile plot for discussion messages.

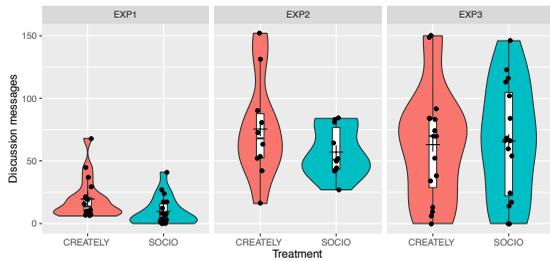

Fig. 9. Violin plot for discussion messages (jitter added to the points).

TABLE 4
DESCRIPTIVE STATISTICS FOR TIME SPENT ON TASKS

| Exp | Treatment | Team | Mean | Std. Dev. | Me- dian |
|---|---|---|---|---|---|
| EXP1 | Creately | 18 | 28.83 | 1.76 | 30.00 |
| EXP1 | SOCIO | 18 | 27.06 | 2.62 | 27.00 |
| EXP2 | Creately | 10 | 27.10 | 2.69 | 28.00 |
| EXP2 | SOCIO | 10 | 25.30 | 2.63 | 25.00 |
| EXP3 | Creately | 16 | 29.19 | 1.42 | 30.00 |
| EXP3 | SOCIO | 16 | 29.19 | 1.56 | 30.00 |

TABLE 5
ANOVA TABLE FOR TIME

| | numDF | denDF | F-value | p-value |
|---|---|---|---|---|
| (Intercept) | 1 | 42 | 15016.244 | <.0001 |
| Sequence | 1 | 40 | 0.010 | 0.9213 |
| **Treatment** | **1** | **42** | **6.187** | **0.0169** |
| Period | 1 | 42 | 0.634 | 0.4305 |
| Experi- ment | 2 | 40 | 11.975 | 0.0001 |

TABLE 6
CONTRAST BETWEEN TREATMENTS FOR TIME

| Contrast | Estimate | SE | df | t-ratio | p-value |
|---|---|---|---|---|---|
| CR-SC | **1.14** | **0.457** | **42** | **2.487** | **0.0169** |

*Fluency.* As Figs. 8 and 9 show, the participants tend to send more messages with Creately than with SOCIO. This can also be corroborated by looking at the descriptive statistics (Table 7). Besides, as shown in the ANOVA table (Table 8), the difference between the number of messages is statistically significant. In particular, **the participants send up to 7.23 more messages with Creately than with SOCIO,** as shown in Table 9. As applies in the case of time, this difference may appear to be small, but it is regarded as a substantial effect in the context of the experimental tasks. Some participants send a single-sentence message containing all the relevant information, whereas others send several smaller messages in a row, all dealing with the same topic. Considering these two different styles of textual communication, experimenters reviewed and counted participant conversations carefully.

We treated a single sentence (containing a complete subject, predicate and object) or an emoji as a message. We considered that a small number of messages meant that the communication effort on the part of participants was smaller. Since both tools support real-time collaboration, users could observe the changes in the class diagram immediately. Also, there were group discussions among users on proper tool use. This type of message accounted for only 11% of the messages. Tool usage messages exhibit a pattern similar to discussion messages. Creately requires more messages about the tool (2.23 messages on average) than SOCIO. This value is statistically significant (p-value = 0.012).

TABLE 7
DESCRIPTIVE STATISTICS FOR DISCUSSION MESSAGES

| Exp | Treatment | Team | Mean | Std. Dev. | Me- dian |
|---|---|---|---|---|---|
| EXP1 | Creately | 18 | 19.56 | 16.30 | 13.50 |
| EXP1 | SOCIO | 18 | 9.61 | 11.51 | 5.00 |
| EXP2 | Creately | 10 | 75.40 | 40.84 | 68.00 |
| EXP2 | SOCIO | 10 | 57.00 | 19.98 | 51.00 |
| EXP3 | Creately | 16 | 63.00 | 45.85 | 70.00 |
| EXP3 | SOCIO | 16 | 65.81 | 46.00 | 66.00 |

We noticed that some teams communicate a lot, whereas others are rather quiet. Our experiments were conducted in different countries, and we found that students in Ecuador tend to be less communicative than Spanish students. This may be due to cultural differences.

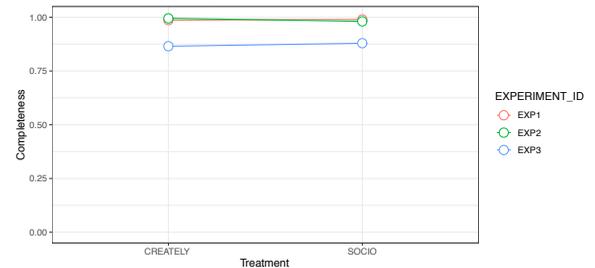

Fig. 10. Profile plot for completeness.

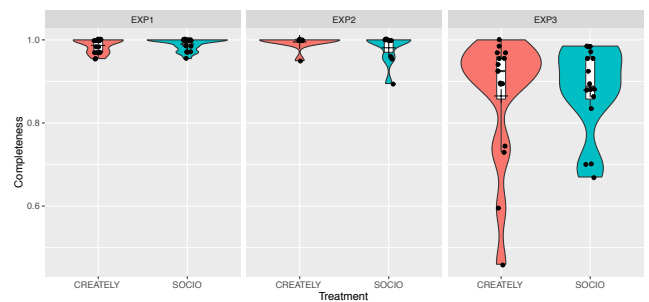

Fig.11. Violin plot for completeness (jitter added to the points).

TABLE 8
ANOVA TABLE FOR DISCUSSION MESSAGES

| | numDF | denDF | F-value | p-value |
|---|---|---|---|---|
| (Intercept) | 1 | 42 | 92.38038 | <.0001 |
| Sequence | 1 | 40 | 0.85428 | 0.3609 |
| **Treatment** | **1** | **42** | **4.11829** | **0.0488** |
| Period | 1 | 42 | 5.69647 | 0.0216 |
| Experi- ment | 2 | 40 | 14.44182 | <.0001 |

TABLE 9
CONTRAST BETWEEN TREATMENTS FOR DISCUSSION MESSAGES

| Contrast | Estimate | SE | df | t-ratio | p-value |
|---|---|---|---|---|---|
| CR-SC | **7.23** | 3.56 | 42 | 2.029 | **0.0488** |

### 5.2.2 Effectiveness

We used the degree of completeness of the tasks to measure effectiveness. Fig. 10 shows the profile plot for the mean of completeness scores of the teams. Fig. 11 shows the violin plot for the completeness scores of the teams. The respective summary of descriptive statistics is shown in Table 10, grouped by experiment and treatment.

TABLE 10
DESCRIPTIVE STATISTICS FOR COMPLETENESS

| Exp | Treatment | Team | Mean | Std. Dev. | Median |
|---|---|---|---|---|---|
| EXP1 | Creately | 18 | 0.99 | 0.02 | 1.00 |
| EXP1 | SOCIO | 18 | 0.99 | 0.01 | 1.00 |
| EXP2 | Creately | 10 | 0.99 | 0.02 | 1.00 |
| EXP2 | SOCIO | 10 | 0.98 | 0.04 | 1.00 |
| EXP3 | Creately | 16 | 0.86 | 0.15 | 0.92 |
| EXP3 | SOCIO | 16 | 0.88 | 0.11 | 0.89 |

As the plots and the descriptive statistics (Figs. 10 and 11 and Table 10) show, completeness appears to be similar for both Creately and SOCIO. Besides, as shown in the ANOVA table (Table 11) and the pairwise contrast between the treatments (Table 12), a negligible –and non-statistically significant– difference in the completeness was observed between Creately and SOCIO (-0.0033). In sum, **Creately and SOCIO appear to perform similarly in terms of completeness.**

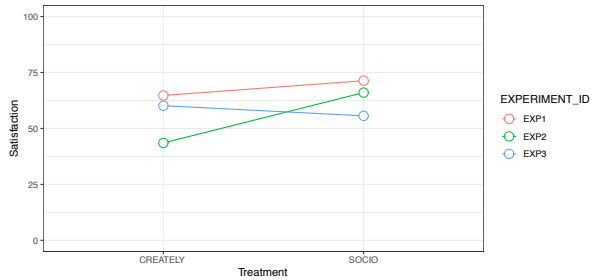

Fig. 12. Profile plot for satisfaction.

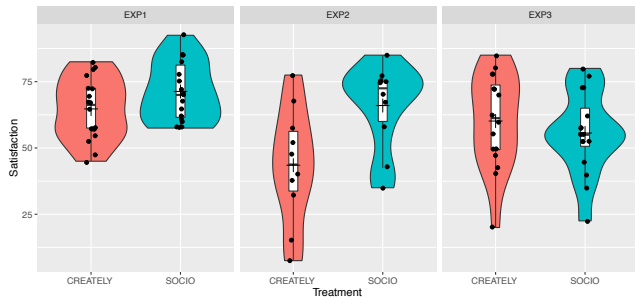

Fig. 13. Violin plot for satisfaction (jitter added to the points).

TABLE 11
ANOVA TABLE FOR COMPLETENESS

| | numDF | denDF | F-value | p-value |
|---|---|---|---|---|
| (Intercept) | 1 | 42 | 8282.362 | <.0001 |
| Sequence | 1 | 40 | 0.775 | 0.3841 |
| Treatment | 1 | 42 | 0.068 | 0.7955 |
| Period | 1 | 42 | 3.076 | 0.0867 |
| Experiment | 2 | 40 | 14.456 | <.0001 |

TABLE 12
CONTRAST BETWEEN TREATMENTS FOR COMPLETENESS

| Contrast | Estimate | SE | df | t-ratio | p-value |
|---|---|---|---|---|---|
| CR-SC | **-0.0033** | 0.0126 | 42 | -0.261 | **0.7955** |

### 5.2.3 Satisfaction

We used the SUS score to quantify satisfaction [19]. Ease of use and learnability are two of the measured sub-characteristics included in SUS questions [49].

Fig. 12 shows the profile plot for the mean SUS scores of the teams. Fig. 13 shows the violin plot for the SUS scores of the teams. The respective summary of descriptive statistics is shown in Table 13, grouped by experiment and treatment.

TABLE 13
DESCRIPTIVE STATISTICS FOR SATISFACTION

| Exp | Treatment | Team | Mean | Std. Dev. | Median |
|---|---|---|---|---|---|
| EXP1 | Creately | 18 | 64.72 | 11.50 | 66.25 |
| EXP1 | SOCIO | 18 | 71.32 | 11.18 | 70.00 |
| EXP2 | Creately | 10 | 43.50 | 21.86 | 43.75 |
| EXP2 | SOCIO | 10 | 66.00 | 16.12 | 72.50 |
| EXP3 | Creately | 16 | 60.16 | 17.78 | 61.25 |
| EXP3 | SOCIO | 16 | 55.62 | 15.51 | 55.00 |

As Figs. 12 and 13 and the descriptive statistics (Table 13) show, the participants appear to be more satisfied with SOCIO than with Creately in two out of three of the experiments (i.e., EXP1 and EXP2). The opposite holds for EXP3, albeit to a smaller extent.

As the ANOVA table (Table 14) and the contrast table (Table 15) show, the difference in satisfaction scores appears to be significant at the 0.1 level. In other words, **the satisfaction scores appear to be higher for participants using SOCIO than for Creately.**

TABLE 14
ANOVA TABLE FOR SATISFACTION

| | numDF | denDF | F-value | p-value |
|---|---|---|---|---|
| (Intercept) | 1 | 42 | 1350.8246 | <.0001 |
| Sequence | 1 | 40 | 0.8612 | 0.3590 |
| **Treatment** | **1** | **42** | **3.4203** | **0.0714** |
| Period | 1 | 42 | 5.4930 | 0.0239 |
| Experiment | 2 | 40 | 5.8296 | 0.0060 |

TABLE 15
CONTRAST BETWEEN TREATMENTS FOR SATISFACTION

| Contrast | Estimate | SE | df | t-ratio | p-value |
|----------|----------|------|----|---------|---------|
| CR-SC | -6.16 | 3.33 | 42 | -1.849 | 0.0714 |

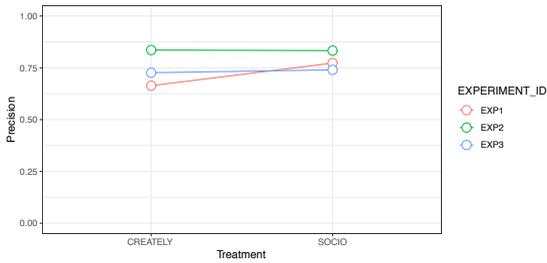

Fig. 14. Profile plot for precision.

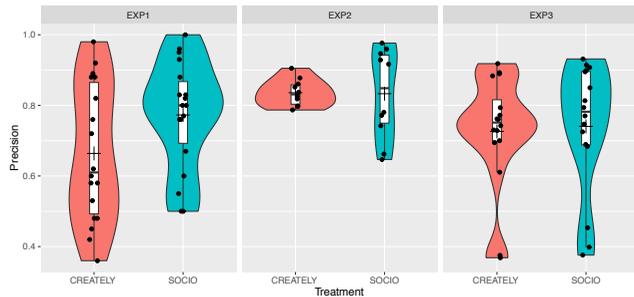

Fig. 15. Violin plot for precision (jitter added to the points).

### 5.2.4 Quality

We analysed the quality of the class diagrams with respect to precision, recall, accuracy, error and perceived success. The profile plots for these metrics are shown in Figs. 14, 16, 18, 20 and 22, respectively, while the violin plots are shown in Figs. 15, 17, 19, 21 and 23. The respective summary of descriptive statistics is shown in Tables 16, 19, 22, 25 and 28.

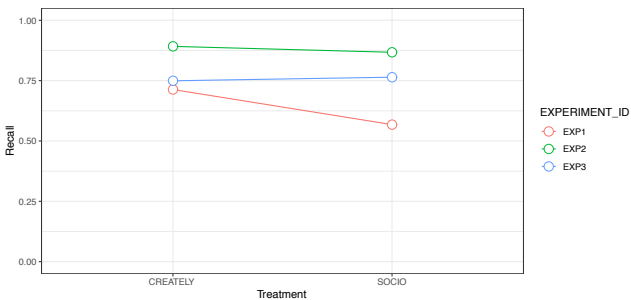

Fig. 16. Profile plot for recall.

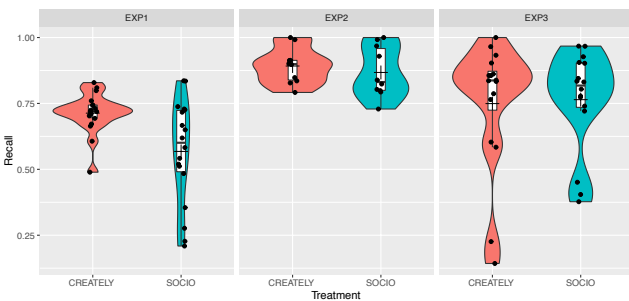

Fig. 17. Violin plot for recall (jitter added to the points).

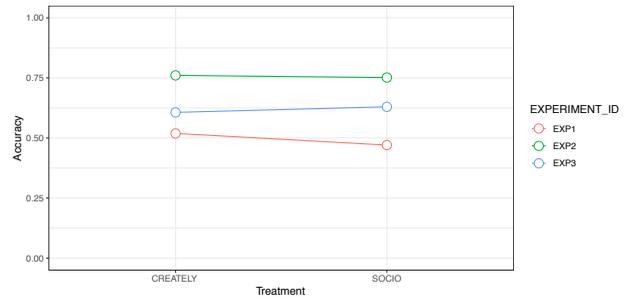

Fig. 18. Profile plot for accuracy.

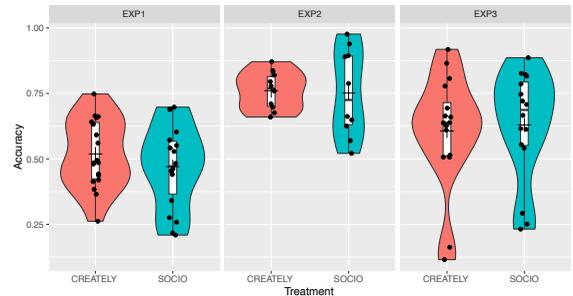

Fig. 19. Violin plot for accuracy (jitter added to the points).

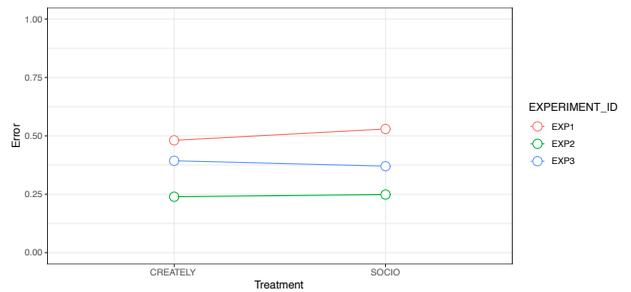

Fig. 20. Profile plot for error.

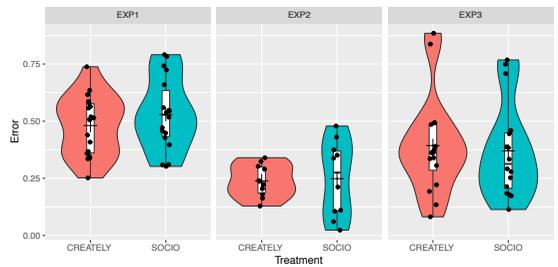

Fig. 21. Violin plot for error (jitter added to the points).

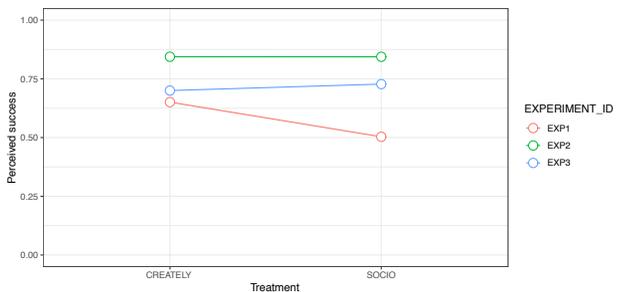

Fig. 22. Profile plot for perceived success.

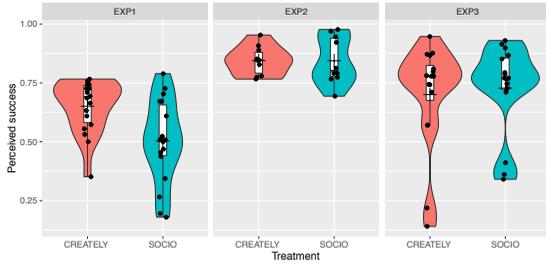

Fig. 23. Violin plot for perceived success (jitter added to the points).

*Precision.* As the plots and the descriptive statistics show, precision appears, on the whole, to be similar for both treatments in most experiments. However, SOCIO outperforms Creately in one experiment (i.e., EXP1). As the ANOVA table (Table 17) shows, the difference in precision between the treatments is statistically significant. In particular, **the participants tend to achieve a higher precision score with SOCIO than with Creately (mean=-0.0491, see Table 18).**

TABLE 16
DESCRIPTIVE STATISTICS FOR PRECISION

| Exp | Treatment | Team | Mean | Std. Dev. | Median |
|-----|-----------|------|------|-----------|--------|
| EXP1 | Creately | 18 | 0.66 | 0.20 | 0.61 |
| EXP1 | SOCIO | 18 | 0.77 | 0.15 | 0.80 |
| EXP2 | Creately | 10 | 0.84 | 0.04 | 0.83 |
| EXP2 | SOCIO | 10 | 0.83 | 0.13 | 0.85 |
| EXP3 | Creately | 16 | 0.73 | 0.16 | 0.75 |
| EXP3 | SOCIO | 16 | 0.74 | 0.18 | 0.78 |

TABLE 17
ANOVA TABLE FOR PRECISION

| | numDF | denDF | F-value | p-value |
|-----|-------|-------|---------|---------|
| (Intercept) | 1 | 42 | 1693.4100 | <.0001 |
| Sequence | 1 | 40 | 0.4214 | 0.5199 |
| **Treatment** | **1** | **42** | **3.9162** | **0.0544** |
| Period | 1 | 42 | 30.4247 | <.0001 |
| Experiment | 2 | 40 | 3.2065 | 0.0511 |

TABLE 18
CONTRAST BETWEEN TREATMENTS FOR PRECISION

| Contrast | Estimate | SE | df | t-ratio | p-value |
|----------|----------|-----|-----|---------|---------|
| CR-SC | **-0.0491** | 0.0248 | 42 | -1.979 | **0.0544** |

*Recall.* As the plots and the descriptive statistics show, both treatments appear to perform similarly in terms of recall in two out of three of the experiments. However, Creately outperforms SOCIO in terms of recall in EXP1.

As the ANOVA table (Table 20) shows, the difference between the treatments for recall is statistically significant (mean=0.0595, see Table 21), where recall is greater for Creately than for SOCIO. Accordingly, **Creately slightly outperforms SOCIO in terms of recall.**

TABLE 19
DESCRIPTIVE STATISTICS FOR RECALL

| Exp | Treatment | Team | Mean | Std. Dev. | Median |
|-----|-----------|------|------|-----------|--------|
| EXP1 | Creately | 18 | 0.71 | 0.08 | 0.72 |
| EXP1 | SOCIO | 18 | 0.57 | 0.20 | 0.60 |
| EXP2 | Creately | 10 | 0.89 | 0.07 | 0.90 |
| EXP2 | SOCIO | 10 | 0.87 | 0.10 | 0.83 |
| EXP3 | Creately | 16 | 0.75 | 0.25 | 0.84 |
| EXP3 | SOCIO | 16 | 0.76 | 0.19 | 0.82 |

TABLE 20
ANOVA TABLE FOR RECALL

| | numDF | denDF | F-value | p-value |
|-----|-------|-------|---------|---------|
| (Intercept) | 1 | 42 | 1234.5463 | <.0001 |
| Sequence | 1 | 40 | 0.2032 | 0.6546 |
| **Treatment** | **1** | **42** | **4.5676** | **0.0384** |
| Period | 1 | 42 | 10.7084 | 0.0021 |
| Experiment | 2 | 40 | 9.7432 | 0.0004 |

TABLE 21
CONTRAST BETWEEN TREATMENTS FOR RECALL

| Contrast | Estimate | SE | df | t-ratio | p-value |
|----------|----------|-----|-----|---------|---------|
| CR-SC | **0.0595** | 0.0279 | 42 | 2.137 | **0.0384** |

*Accuracy.* As the plots, descriptive statistics, ANOVA table (Table 23) and contrast (Table 24) show, **Creately and SOCIO both tend to perform similarly in terms of accuracy.**
*Error.* As the plots, descriptive statistics, ANOVA table (Table 26) and contrast (Table 27) show, **Creately and SOCIO both tend to return a similar number of errors.**

TABLE 22
DESCRIPTIVE STATISTICS FOR ACCURACY

| Exp | Treatment | Team | Mean | Std. Dev. | Median |
|-----|-----------|------|------|-----------|--------|
| EXP1 | Creately | 18 | 0.52 | 0.13 | 0.49 |
| EXP1 | SOCIO | 18 | 0.47 | 0.16 | 0.47 |
| EXP2 | Creately | 10 | 0.76 | 0.07 | 0.77 |
| EXP2 | SOCIO | 10 | 0.75 | 0.17 | 0.73 |
| EXP3 | Creately | 16 | 0.61 | 0.22 | 0.64 |
| EXP3 | SOCIO | 16 | 0.63 | 0.21 | 0.69 |

TABLE 23
ANOVA TABLE FOR ACCURACY

| | numDF | denDF | F-value | p-value |
|-----|-------|-------|---------|---------|
| (Intercept) | 1 | 42 | 864.8574 | <.0001 |
| Sequence | 1 | 40 | 0.2964 | 0.5892 |
| Treatment | 1 | 42 | 0.3241 | 0.5722 |
| Period | 1 | 42 | 34.5075 | <.0001 |
| Experiment | 2 | 40 | 12.2809 | 0.0001 |

TABLE 24
CONTRAST BETWEEN TREATMENTS FOR ACCURACY

| Con-trast | Estimate | SE | df | t-ratio | p-value |
|---|---|---|---|---|---|
| CR-SC | 2 | 40 | 12.2809 | 0.0001 | 0.5722 |

TABLE 25
DESCRIPTIVE STATISTICS FOR ERROR

| Exp | Treatment | Team | Mean | Std. Dev. | Me-dian |
|---|---|---|---|---|---|
| EXP1 | Creately | 18 | 0.48 | 0.13 | 0.51 |
| EXP1 | SOCIO | 18 | 0.53 | 0.16 | 0.53 |
| EXP2 | Creately | 10 | 0.24 | 0.07 | 0.23 |
| EXP2 | SOCIO | 10 | 0.25 | 0.17 | 0.27 |
| EXP3 | Creately | 16 | 0.39 | 0.22 | 0.36 |
| EXP3 | SOCIO | 16 | 0.37 | 0.21 | 0.31 |

TABLE 26
ANOVA TABLE FOR ERROR

| | numDF | denDF | F-value | p-value |
|---|---|---|---|---|
| (Intercept) | 1 | 42 | 387.6832 | <.0001 |
| Sequence | 1 | 40 | 0.2964 | 0.5892 |
| Treatment | 1 | 42 | 0.3241 | 0.5722 |
| Period | 1 | 42 | 34.5075 | <.0001 |
| Experiment | 2 | 40 | 12.2809 | 0.0001 |

TABLE 27
CONTRAST BETWEEN TREATMENTS FOR ERROR

| Contrast | Estimate | SE | df | t-ratio | p-value |
|---|---|---|---|---|---|
| CR-SC | -0.0134 | 0.0236 | 42 | -0.569 | 0.5722 |

*Perceived success.* As shown in the ANOVA table (Table 29) and by the contrast (Table 30), the difference in perceived success is significant at the 0.1 level, where Creately outperforms SOCIO.

TABLE 28
DESCRIPTIVE STATISTICS FOR PERCEIVED SUCCESS

| Exp | Treatment | Team | Mean | Std. Dev. | Me-dian |
|---|---|---|---|---|---|
| EXP1 | Creately | 18 | 0.65 | 0.11 | 0.69 |
| EXP1 | SOCIO | 18 | 0.50 | 0.18 | 0.51 |
| EXP2 | Creately | 10 | 0.84 | 0.06 | 0.85 |
| EXP2 | SOCIO | 10 | 0.84 | 0.10 | 0.80 |
| EXP3 | Creately | 16 | 0.70 | 0.23 | 0.78 |
| EXP3 | SOCIO | 16 | 0.73 | 0.19 | 0.77 |

TABLE 29
ANOVA TABLE FOR PERCEIVED SUCCESS

| | numDF | denDF | F-value | p-value |
|---|---|---|---|---|
| (Intercept) | 1 | 42 | 1139.2168 | <.0001 |
| Sequence | 1 | 40 | 0.1317 | 0.7186 |
| **Treatment** | **1** | **42** | **3.7321** | **0.0601** |
| Period | 1 | 42 | 15.3435 | 0.003 |
| Experiment | 2 | 40 | 13.0200 | <.0001 |

TABLE 30
CONTRAST BETWEEN TREATMENTS FOR PERCEIVED SUCCESS

| Contrast | Estimate | SE | df | t-ratio | p-value |
|---|---|---|---|---|---|
| CR-SC | 0.0503 | 0.026 | 42 | 1.932 | **0.0601** |

## 5.3 Discussion of Analysis

Table 31 shows a summary of the results at family level. The student participants spend more time and send a larger number of messages with Creately than with SO-CIO. In other words, **SOCIO outperforms Creately in terms of efficiency**. A similar observation holds for satisfaction. Overall, we can conclude that **the student participants appear to be more satisfied with SOCIO than with Creately**. Finally, **SOCIO outperforms Creately in terms of precision**.

However, **Creately outperforms SOCIO in terms of recall and perceived success**. In other words, the student participants using SOCIO appear to have developed a small number of classes, most of which are within the ideal solution. On the other hand, **the student participants using Creately created more classes.** Therefore, they had **the impression that they were more successful with Creately than with SOCIO.**

## 6 THREATS TO VALIDITY

In this section, we discuss the main threats to the validity of our SMS and the family of experiments. First of all, we discuss the threats to our SMS.

When conducting a SMS, the search terms should identify as many relevant papers as possible. We piloted different terms and search strings along the way to reduce the risk of missing relevant primary studies. Also, we used synonyms of the terms taken from papers related to the objective of the research. The selection of primary studies within the scope of the research is crucial for achieving relevant conclusions.

The protocol included the definition of explicit exclusion criteria to minimize subjectivity and prevent the omission of valid articles. Exclusion criteria were applied by two researchers to select valid primary studies. We have uploaded this SMS protocol to http://dx.doi.org/10.6084/m9.figshare.19142012. The threats to the validity of our family of experiments are discussed below.

### 6.1 Statistical Conclusion Validity

Threats to conclusion validity may materialize due to the presence of small sample sizes (increasing the potential for inaccurate results). We tried to mitigate this shortcoming by replicating the baseline experiment as closely as possible twice and aggregating the results of all the experiments together. Although this number of subjects may not suffice to get accurate results for all the metrics, the increase in the sample size of the results from 18 to 44 was greater than the minimum sample size required for all variables except two (discussion messages and satisfaction). It was not necessary to repeat the experiment with more subjects for these two variables because the results were significant. These rules out any type II errors.

TABLE 31
SUMMARY OF RESULTS AT FAMILY LEVEL

| Time | N. Discussion | Completeness | Satisfaction | Precision | Recall | Accuracy | Error | Success |
|------|---------------|--------------|--------------|-----------|--------|----------|-------|---------|
| **CR>SC** | **CR>SC** | - | *CR<SC* | *CR<SC* | **CR>SC** | - | - | **CR>SC** |

However, the above should be viewed with caution. The experimental design contains nine dependent variables. Each variable was analysed using a different LMM, which inflates the type I error. To prevent inflation, we have two options: 1) use a multivariate LMM (which is seldom applied in SE), or 2) correct the alpha level used to claim that a result is significant. Using the simplest procedure (Bonferroni), the alpha level should have dropped from 0.05 to 0.0056.

The problem of reducing the alpha value (to prevent type I error) is that it also reduces statistical power, increasing type II error. Without the Bonferroni correction, the LMM-wise type-I error is $P(B(0.05, 9)>= 1) = 37\%$, that is, we have a 37% increase in the error rate for all 9 LMMs. Using Bonferroni, the type II error is over 60% for four variables, and increases significantly for the remainder. We are talking about increases of over 40% per variable (not variable-wise). Therefore, the reasonable option is to use an alpha value of 0.05 and assume that there is a 37% probability of a type I error rather than risking enormous type II errors.

Threats to conclusion validity may also emerge due to the use of inappropriate analysis procedures. With the aim of mitigating this shortcoming, we relied upon parametric statistical tests (i.e., LMM [59]) to analyse the data of our family of experiments. LMMs are typically used in mature experimental disciplines such as medicine for analysing multi-site experiments [13]. We ensured the robustness of the results by meta-analysing the data with a one-stage IPD model and accounting for an extra factor (i.e., experiment) in order to model the difference between the results across the experiments [13], [61], [62].

The data are slightly heterogeneous, which is appreciable for the TOOL_USAGE_MESSAGES and PRECISION response variables only. To evaluate the impact of heterogeneity on the analyses, we conducted a generalized least squares analysis [63] to check the consistency of the conclusions based on the mixed model fitted using REML. As the results for both models were similar (with respect to effect size and statistical significance), we interpreted the results using the parsimonious model (that is, the linear mixed model). In order to ensure the transparency of the results, we provide the original data and statistical analyses carried out in the supplementary materials. In the spirit of open science, we have uploaded the experimental data and the materials of the experiments to http://dx.doi.org/10.6084/m9.figshare.19142012.

## 6.2 Internal Validity

Threats to internal validity may also materialize because the participants learn by practice, which boosts participant performance in the second session (when developing the second-class diagram). We tried to minimize this threat by using a cross-over experimental design (so both treatments benefit equally from this circumstance).

Another threat is tiredness or boredom since the sessions are one-and-a-half hours long. Participants may not be very motivated since it is a voluntary experiment that does not have any impact on their grades. Still, we think that these analyses may serve to foster further research to evaluate chatbot or modelling tool usability.

Carryover is another of the threats to experiments with a crossover design. Carryover occurs when applying a treatment before the effect of the previous treatment has disappeared. As a result, if the first-applied treatments improve (reduce) the effectiveness of the treatments applied later, the first-applied treatments may appear to be comparatively less (more) effective. Also, in experiments with a crossover design, where the number of treatments and periods is the same, carryover may be confounded with the treatment-period interaction and sequence effects. Therefore, it is impossible to discern which of the three effects is really at work [48].

## 6.3 Construct Validity

Although our tasks do not require complex or high-level English communication, we rate students' English level to filter out participants with low English proficiency because the Creately tool menu options and SOCIO commands are in English. In order to ensure the smooth progress of the experiment, we translated the experimental materials into the participants' native language (Spanish) so that they did not have to spend time and mental effort on language translation. We acknowledge that, although the experimental material was translated into the participants' native language to make them feel comfortable, the self-assessment questions may not accurately capture their background [64], [65]. Thus, the use of questionnaires may have biased the results of the satisfaction response variable. However, this approach has been used in other studies to measure satisfaction.

Participant subjects exchange messages for different purposes, that is, they do not all refer to tool use. For example, they include complaints, jokes, questions for experimenters, etc. Therefore, we identified messages referred strictly to tools as a subset of discussion messages, which have been accounted for by the tool usage messages variable. This variable was measured manually, on which ground our criteria may differ from other researchers'.

## 6.4 External Validity

As usual in SE experiments [66], we had to rely on toy tasks and students to evaluate the performance of the treatments. This limits the external validity of the results. Thus, we acknowledge that our findings are limited to academia and may not be generalizable to industry (despite the fact that student participants were acquainted with how to design class diagrams). Besides, the results are limited to structural conceptual diagrams like class diagrams, the results cannot be extended to other model and diagram types.

## 7 DISCUSSION OF RESULTS

Two other small-scale evaluations [4], [21], mentioned in Section 3, assessed SOCIO chatbot usability. In the first evaluation, it achieved a positive satisfaction result (74%) [4], although they found that natural language interpretation required improvement. The results of the second evaluation [21] found that the consensus mechanism was considered useful for large groups. Our family of experiments is the first and only research to evaluate the usability of the SOCIO chatbot comprehensively with regard to effectiveness, efficiency and satisfaction. Additionally, while SOCIO chatbot usability has been evaluated previously, the number of subjects was smaller than provided by this family of experiments, and it was not compared with other tools. Besides, there is, to the best of our knowledge, no other chatbot offering similar functions or services to SOCIO. In view of this, we identified the need to conduct an experiment on the comparative usability of the collaborative modelling SOCIO chatbot in a family of experiments in academic settings. This is the original contribution of this research.

Regarding the effectiveness and efficiency results, it appears that 44.1% of the subjects tend to spend as long as possible on completing and/or improving their class diagrams, while 55.9% completed the class diagram in the shortest possible time, that is, they finished their assignment before the 30-minute time limit. The above 44.1% of subjects finished the class diagram that they were set as a task on the verge of the 30-minute time limit. If they had been given more time, they would have used it. In this case, the time taken would have been longer on average. However, we decided to establish a time limit to be able to measure other variables, such as task completion rate. We would not have been able to measure this effectiveness variable if all the experimental subjects had had the option of completing the task.

Regarding satisfaction, we also extended the SUS questionnaire with four open-ended questions (concerning positive and negative aspects of the two tools, suggestions and user preferences) in order to gather definite satisfaction-related opinions from subjects.

In response to the open-ended questions, many participants remarked that they found both tools to be satisfactory in terms of responsiveness, ease of use, and collaboration capabilities. Besides, Creately was praised for its friendly interface, whereas SOCIO was more fun to use.

However, the participants also registered some complaints. We have translated their comments below. With regard to the SOCIO chatbot, 24.7% of participants complained about the language limitations of the chatbot, for example, one participant said that the chatbot only recognized one language.

Of the participants, 24.7% complained that the SOCIO chatbot commands, especially the /undo command, are not easy to learn. For example, one participant stated that the /undo command undid the last change made by all group members, whereas it should only delete the change made by the person using the undo command. Another participant complained that the command-based use of the tool was complicated.

Of the sample, 16.4% complained about the SOCIO chatbot help web page and responses that offer help, for example, one participant criticized the fact that the documentation was not very clear and was not divided by topics such as 'Relationships', 'Attribute Creation', 'Attribute Removal', etc.

On the other hand, the biggest problems with Creately were related to real-time collaboration. Of the participants, 24.7% claimed that Creately produced some errors when loading on some users' computers, and one participant stated that it was slow for poor internet connections and did not run well. Another participant said that he was unable to participate in collaborative work, as the following error message was displayed: 'We have logged an error in Creately. You can continue working with Creately, but we recommend that you save your work to avoid losses.' However, he remarked that he could not even place objects on the workspace canvas, let alone save his work. He claimed that he was not the only one with this problem.

With a view to improving understandability, we believe it is necessary to single out the results that differentiate this research from the baseline experiment [19]. The baseline experiment confirms that SOCIO requires less time, generates fewer discussion messages, and has more precision, less recall, and less perceived success than Creately. The other variables (tool usage messages, completeness, satisfaction, accuracy, and error) yielded non-significant results.

The family of experiments increases the precision of the results of the baseline experiment thanks to the bigger sample size. The additional experiments confirm most of the previous results, although some are challenged. In particular, the tool usage messages are now significant, indicating that the groups using Creately need more information about how the tool works than SOCIO teams.

The cumulative meta-analysis (see Appendix C) also shows that, generally, the 95% confidence intervals get narrower, meaning that the pooled results are more and more precise as the results pile up. Compared to single experiments, a family provides greater confidence that the results are correct and more accurate estimates of the treatment effects. Such confidence is not limited to significant results. Non-significant results can also be confidently interpreted. As reported in Appendix B, 50 teams are needed to detect small-to-medium effect sizes with an 80% power. We have assembled a cohort of 44 teams, quite close to the required sample size. The most likely conclusion is that non-significant variables produce small or very small effect sizes. Therefore, these variables are not of practical interest from the viewpoint of chatbot usability.

In fact, the baseline experiment [19] was quite precise. The results of the first experiment match the final meta-analysis for eight out of ten response variables (including the response variable that refers to the discussion messages related exclusively to tool usage). This was somehow to be expected. For most response variables, the required sample size to achieve 80% power (see Appendix B) lies between 10 and 35 teams per sequence. The baseline experiment has 18, which lies in between these figures.

As mentioned before, our family of experiments increases the precision of the results of the baseline experiment thanks to the bigger sample size. Thus, it is possible to clarify the required evidence-based SOCIO chatbot improvements. Currently, work is underway to develop different updated versions of the SOCIO chatbot:

1) Provide alternative context-sensitive help when the SOCIO chatbot has difficulties understanding the user.

2) Add functionalities requested by users: users will be able to delete any elements that they like by clicking the buttons underneath, and users will be able to choose how many steps to cancel or redo at a time instead of deleting or redoing one by one.

3) Provide an option to select class diagram appearance.

4) Update and supplement the help page for all three versions in more than one language.

## 8 CONCLUSIONS AND FUTURE WORK

SOCIO chatbot usability was evaluated through comparison with the Creately web-based application. A total of three experiments were conducted in order to form a family of experiments and provide a larger data set [67]. We used identical tasks and response variable operationalizations across all the experiments. A total of 132 student participants were recruited in our experiments, forming 44 teams. After the experiments, we reported joint results and assessed the potential effects of the experimental changes with meta-analysis [60].

With our family we observed that the chatbot gets better scores in terms of **efficiency**. Regarding **effectiveness**, similar results were reported with both treatments at the family level. For **satisfaction**, SOCIO has better SUS scores. Finally, considering **the quality of the class diagrams**, precision scores are higher for SOCIO, while Creately appears to be better in terms of recall and correct variables.

In conclusion, we fail to reject the null hypothesis H.2.0 and reject the null hypotheses H.1.0, H.3.0 and H.4.0.

> The student participants also appear to create fewer classes with SOCIO, and most elements within those classes are correct if assessed against the ideal solution. However, the student participants created more classes with Creately and achieved better results than with SOCIO.

Although the family of experiments reinforced the previous result, some differences should be pointed out. Regarding efficiency, the difference between the two tools is narrower in the family of experiments than in the baseline experiment. Specifically, the time difference used was reduced from 1 minute and 47 seconds to 1 minute and 8 seconds, and the difference in messages sent by users was reduced from 10 to about 7.23. Regarding satisfaction, EXP3 shows the opposite result to the baseline experiment and EXP2: users found Creately to be slightly more satisfactory than SOCIO.

Our experiments provide developers with a different angle on the usability of the SOCIO chatbot and the Creately online tool. We compared both tools to determine which one helps teams to be more effective and efficient at generating class diagrams with a view to improving SOCIO chatbot usability. We did not set out to determine how three-member teams can build better class diagrams collaboratively. Additionally, this research contributes to the body of empirical evidence on the chatbot and collaborative modelling tool in terms of usability. It tackles a chatbot-based approach to software engineering that may reduce the gap between requirements engineering and modelling. We report statistically significant differences with a medium effect size.

In view of the small effect and weight of the cumulative results with respect to the final result so far, especially as regards effectiveness and satisfaction, it would not appear to be worthwhile repeating the same experiment. Therefore, further studies will continue along the following lines: (i) conduct further replications with native English-speaking subjects, (ii) reframe the experimental design by extending time limits for task completion, and (iii) carry out more replications to compare the usability of the original and updated versions of the SOCIO chatbot. We only analysed the results for three-member teams. In this respect, the research is limited, and larger sizes such as four- or five-member teams explored in previous research [21] and possibly also the performance of single modellers with and without the SOCIO chatbot should be examined. Finally, we cannot assume that the results are independent of the characteristics of the experimental subjects. This is an issue that should be addressed in future work, which should analyse the impact of moderator variables, like subject sociodemographic characteristics, on the results of the family of experiments.

## ACKNOWLEDGMENT

This research was funded by Spanish Ministry of Science, Innovation and Universities research grant PGC2018-097265-B-I00 and MASSIVE project (RTI2018-095255-B-I00). This research was also supported by the Madrid Region R&D programme (project FORTE, P2018/TCS-4314).

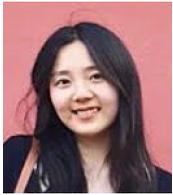

**Ranci Ren** received her MS in ICT Research and Innovation from the Universidad Autónoma de Madrid (UAM), Spain, in 2019. She is currently working toward her PhD in software engineering at UAM. Her main research interests include experimental software engineering, human-computer interaction, and chatbots. She is a member of the ACM.

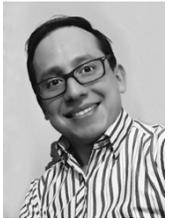

**John W. Castro** received his PhD from the Universidad Autónoma de Madrid in 2015. He received his MS in Computer Science and Telecommunications, specializing in Advanced Software Development, from the Universidad Autónoma de Madrid in 2009. He has fifteen years of experience in the area of software system development. He is currently assistant professor at the Universidad de Atacama (Chile). He worked as a research assistant at the Universidad Politécnica de Madrid. His research interests include software engineering, software development process, and the integration of usability in the software development process.

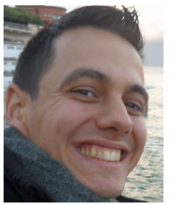

**Adrián Santos** received his MSc in Software and Systems and MSc in Software Project Management from the Universidad Politécnica de Madrid, Spain, and his MSc in IT Auditing, Security and Government from the Universidad Autónoma de Madrid, Spain. He received his PhD in Software Engineering from the University of Oulu, Finland. He is currently working as a software engineer in industry.

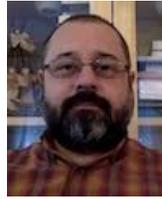

**Oscar Dieste** received his BS and MS in Computing from the Universidad da Coruña and his PhD from the Universidad de Castilla La Mancha. He is a researcher with the UPM's School of Computer Engineering. He was previously with the University of Colorado at Colorado Springs (as a Fulbright scholar), the Universidad Complutense de Madrid, and Universidad Alfonso X el Sabio. His research interests include empirical software engineering and requirements engineering.

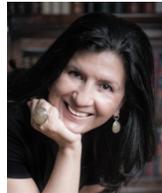

**Silvia T. Acuña** received her PhD from the Universidad Politécnica de Madrid in 2002. She is currently an associate professor of software engineering at Universidad Autónoma de Madrid's Computer Science Department. Her research interests include experimental software engineering, software usability, software process modelling, and software team building. She co-authored A Software Process Model Handbook for Incorporating People's Capabilities (Springer, 2005), and edited Software Process Modeling (Springer, 2005) and New Trends in Software Process Modeling (World Scientific, 2006). She is the deputy conference co-chair on organizing committee of ICSE 2021. She is a member of the IEEE Computer Society and a member of the ACM.

## APPENDIX A CHATBOT USABILITY EXPERIMENTATION

In systematic mapping study (SMS), we reviewed a total of 25 primary studies that included experiments on chatbot usability (see Section 3 Related Work). As shown in Table A1, we summarized these 25 experiments, including information like the goal, article research questions (ARQ), number of experiments in the family, sample sizes, and type of subjects.

TABLE A1
Experiments on Chatbot Usability

| Ref. | Goal | Research Questions | Size | Sample Sizes | Type of Subjects |
|---|---|---|---|---|---|
| [19] | Evaluate the usability of the SOCIO chatbot through comparison with the Creately web tool with respect to user effectiveness, efficiency and satisfaction and the quality of the resulting class diagrams. | (ARQ1.1) Compared to Creately, does the use of SOCIO positively affect user efficiency, effectiveness and satisfaction when building class diagrams and class diagram quality? | 1 | 54 | Graduates and undergraduates |
| [24] | Ascertain whether the relatively inexpensive approach of using real-time head pose measurements as a proxy for user attention is a suitable alternative to using a wake-up word. | (ARQ2.1) Compared to the chatbot version that requires a wake-up word to signify that users are addressing the assistant, is the advanced chatbot version that requires a head pose more usable and likable? | 1 | 8 | Students |
| [25] | Explore how to communicate service offers as part of chatbot interaction, and user preferences for such service offers. | (ARQ3.1) How can service offers be communicated to users during conversational interactions at different levels of proactivity, and what are the user preferences for such offers? | 1 | 17 | Not defined |
| [26] | Evaluate TV viewers' user experience (UX) of interactions assisted by a conversational agent (CA) compared with the remote-control unit (RCU), while watching TV in terms of pragmatic quality (PQ), hedonic quality (HQ) and attractiveness (ATT). | (ARQ4.1) How does the TV viewers' UX of CA- and RCU-assisted interactions compare in terms of PQ, HQ, and ATT? | 1 | 42 | Students |
| [27] | Investigate system accuracy and interaction efficiency comparing different interaction modes (completely natural language, completely button, and mixed). | (ARQ5.1) To what extent can conversational interfaces support the music recommendation? (ARQ5.2) What is the best conversational interface in terms of the cost of interaction? (ARQ5.3) What is the best conversational interface in terms of recommendation accuracy? (ARQ5.4) Is the disambiguation step particularly strenuous for the user of a conversational music recommender system? | 1 | 110 | Students |
| [28] | Evaluate the efficacy of a new innovative approach for collecting data in health care, namely a virtual conversational agent (VCA) interface, compared with the standard interface for collecting family health histories (FHx). | (ARQ6.1) Compared to the standard interface, does the VCA interface positively affect the interface workload, usability, preference, and satisfaction when collecting FHx? | 1 | 15 | Residents |

TABLE A1
EXPERIMENTS ON CHATBOT USABILITY (CONT'D)

| Ref. | Goal | Research Questions | Size | Sample Sizes | Type of Subjects |
|------|------|--------------------|------|--------------|-------------------|
| [29] | Analyse the perceived usability of the Amazon Echo Dot (3rd generation), Apple HomePod, and Google Nest Mini speech-only contexts, compared with the GUI paradigm. | (ARQ7.1) How does the perceived usability of Amazon Echo Dot (3rd generation), Apple HomePod, and Google Nest Mini speech-only contexts compare with the GUI paradigm? | 1 | 61 | Not defined |
| [30] | Analyse Voice Usability Scale (VUS) usability of the Amazon Echo Dot (3rd generation) speech-only contexts. | (ARQ8.1) How good is the VUS usability of Amazon Echo Dot (3rd generation) speech-only contexts? | 1 | 61 | Not defined |
| [31] | Evaluate ConveRSE's ability to adapt to an interface based on a social humanoid chatbot in terms of both recommendation accuracy and user experience by comparing the smartphone- and robot-based interface. | (ARQ9.1) Can chatbot ConveRSE be implemented through a social robot without losing performance in terms of recommendation accuracy and user experience compared to a chatbot-based interface? | 1 | 20 | Not defined |
| [32] | Compare the conversational search user interface (chatbot) of a medical resource centre database with its graphical search user interface in terms of user engagement and usability. | (ARQ10.1) How does a conversational search interface compare to a graphical search user interface in terms of user engagement and usability? | 1 | 10 | Not defined |
| [33] | Measure and compare the Convey chatbot and the default chatbot in terms of the time taken to complete the task, the total number of words input by the user, and the total number of user actions (they browsed and zoomed in on a particular product). | (ARQ11.1) Compared with the default chatbot, does the Convey chatbot perform better in terms of the time taken to complete the task, the total number of words input by the user, and the total number of the user actions (they browsed and zoomed in on a particular product)? | 1 | 16 | Company employees |
| [34] | Examine usability in terms of the effects of continuous conversation and task complexity on interaction with an AI-infused conversational agent in a simulated smart home environment. | (ARQ12.1) Does the continuous conversation (on vs. off) and task complexity (simple vs. complex) affect the usability of Tianmao jingling chatbot? | 1 | 18 | Company employees |
| [35] | Evaluate user interaction with the chatbot ROB against the original paper version of Adult ADHD Self-Report Scale (ASRS) for screening for symptoms associated with attention deficit hyperactivity disorder. | (ARQ13.1) Compared to the original paper version of ASRS, does the use of chatbot ROB positively affect user interaction quality with respect to the participants' scores on the ASRS and the time for completion? (ARQ13.2) What is the difference in the user experience between the two versions of ASRS? | 1 | 11 | Not defined |
| [36] | Compared with a chatbot communicating in modern English, evaluate how a Shakespearean language style applied to e-commerce chatbots affects customer experience and their attitude towards a presented product. | (ARQ14.1) How does adding language style to an e-commerce chatbot affect user satisfaction, user interest in a product, perceived product value and user engagement? | 1 | 169 | Not defined |

TABLE A1
EXPERIMENTS ON CHATBOT USABILITY (CONT'D)

| Ref. | Goal | Research Questions | Size | Sample Sizes | Type of Subjects |
|---|---|---|---|---|---|
| [37] | Understand the usability of the FarmChat system by comparing two chatbot interfaces (Audio-only FarmChat and Audio+Text Farm-Chat) with differing interaction modalities. | (ARQ15.1) How acceptable is FarmChat as an information system to satisfy farmers' information needs? (ARQ15.2) How usable is FarmChat for interacting with conversational interfaces? (ARQ15.3) What is the preference between the two variants of conversational interfaces –Audio+Text versus Audio-only– and how does it differ for different user populations? | 1 | 34 | Farmers |
| [38] | Investigate the usability of a short-term mobile-based interactive chatbot Todaki in alleviating attention deficit symptoms. | (ARQ16.1) Does the chatbot Todaki report comparable acceptability compared to the book? | 1 | 46 | Patients |
| [39] | Compare rule-based and natural language processing-based chatbots in terms of usefulness, usability, searchability, reliability and attractiveness. | (ARQ17.1) Compared to rule-based chatbot Talkjipsa, does the natural language processing-based chatbot Samantha perform better in terms of usefulness, usability, searchability, reliability and attractiveness? | 1 | 79 | Students |
| [40] | Study the differences in system satisfaction between a chatbot system and a website system and what factors determine satisfaction based on self-determination theory. | (ARQ18.1) What factors affect system satisfaction in a chatbot system? (ARQ18.2) Does the level of system satisfaction differ between a website system and a chatbot system? | 1 | - | Students |
| [41] | Evaluate, compared with baseline conditions, the usability of the situation-aware adaptation of the chatbot with qualitative feedback during the real-world experiment. | (ARQ19.1) Compared with baseline conditions, does the chatbot with situation-aware adaptation have a positive effect with respect to usability? | 1 | 12 | Experienced users |
| [42] | Assess the usability of a mentoring conversational agent comparing mobile interfaces (Twitter and SMS) with each other and against a web-based embodied conversational agent (ECA). | (ARQ20.1) How usable are the mobile conversational agent interfaces compared to the web-based ECA interface? | 2 | 35 | Students |
| [43] | Explore the usability and acceptability of chatbot DynamicDuo in both controlled laboratory-based studies and real-world environments. | (ARQ21.1) Will presenters accept a chatbot as a co-presenter for scientific presentations? | 2 | 12+10 | Students and professionals |
| [44] | Measure perinatal women's and their partners' perceptions of the utilitarian and hedonic value of medical chatbot Dr. Joy experience. | (ARQ22.1) Does the chatbot Dr. Joy produce both utilitarian and hedonic values during the 7-day contextual usability testing period? | 1 | 15 | People in pregnancy preparation or different pregnancy stages were enrolled |

TABLE A1
EXPERIMENTS ON CHATBOT USABILITY (CONT'D)

| Ref. | Goal | Research Questions | Size | Sample Sizes | Type of Subjects |
|------|------|--------------------|------|-------------|------------------|
| [45] | Assess: 1) the effect of perceived similarity between the MyPAL robot and an avatar on children's friendship toward the avatar, and 2) the effect of this friendship on the usability of, and children's motivation to play with, a self-management application containing the avatar. | (ARQ23.1) How does the MyPAL app perform on similarity, friendship, motivation, and usability? (ARQ23.2) What are the relationships between similarity, friendship, motivation, and usability of the MyPAL app performance? | 1 | 21 | Children |
| [46] | Examine how user understanding affects perceptions and experiences of using a CA, specifically Apple Siri. | (ARQ24.1) To what extent does the personal experience of using the CA and the technical knowledge about the CA's system model affect how people feel about the CA? | 1 | 41 | Experienced and inexperienced users, users with and without technical knowledge |
| [47] | Transcribe the real conditions of interactions with a professional virtual agent to capture as accurately as possible the perceptions and usage behaviours of real users. | (ARQ25.1) How would the expression of intimate behaviours by the chatbot impact the users' perception of virtual intimacy, social presence, and user experience in a real-world situation? | 1 | 60 | Visitors to the tourist information office |

## APPENDIX B   ESTIMATION OF THE (A-PRIORI) STATISTICAL POWER

We conducted a power analysis using a Monte-Carlo simulation. The code is available in the reproduction package (see /Data Analysis/03-Power calculation.R). The Monte-Carlo simulation requires the values of the within-subjects and between-subjects variances. We used the variances obtained in the first experiment of the family (EXP1) for this purpose. The power was estimated for the sample sizes (per sequence): 5, 10, 15, 20, 25, 30, 35, 40, 45, 50, 55, 60, 65, 70, 75, 80, 85, 90, 95, 100. For each sample size, 500 iterations were performed.

For each response variable, we set the minimum effect size (in natural units) that we aim to detect for the treatment (SOCIO vs. CREATELY). These effect sizes are rather small, i.e., we want the family of experiments to be quite sensitive. When translated to standardized effect sizes (Cohen's d, because Hedges' g is irrelevant for power estimation), we found that all standardized effects were small to medium. The details are shown in Table B1. The last column reports the required sample size to achieve 80% power for each response variable. Note that we added the TOOL_USAGE_MESSAGES response variable (which refers exclusively to the discussion messages related to tool use). The required sample size for the overall experiment is the maximum of the per-response variable required sample size, that is, **50 teams per sequence, 100 teams in total**.

TABLE B1
EFFECT SIZES FOR EACH RESPONSE VARIABLE

| Response variable | Within-subjects variance | Between-subjects variance | Relevant effect (natural measure) | Relevant effect (standardized effect size) | Sample size to achieve 80% power (per sequence) |
|-------------------|--------------------------|---------------------------|-----------------------------------|--------------------------------------------|-------------------------------------------------|
| TIME | 0.3 | 2.09 | 1 minute | 0.48 | 25 |
| DISCUSSION_MESSAGES | 28.48 | 16.70 | 5 messages | 0.15 | 50 |
| TOOL_USAGE_MESSAGES | 1.88 | 4.32 | 2 messages | 0.42 | 25 |
| COMPLETENESS | 0.06 | 0.06 | 0.05 | 0.13 | 10 |
| SATISFACTION | 4.66 | 11.59 | 5 | 0.40 | 45 |
| PRECISION | 0.11 | 0.12 | 0.05 | 0.31 | 30 |
| RECALL | 0.08 | 0.14 | 0.05 | 0.30 | 35 |
| ACCURACY | 0.11 | 0.11 | 0.05 | 0.32 | 25 |
| ERROR | 0.11 | 0.11 | 0.05 | 0.32 | 25 |
| PERCEIVED_SUCCESS | 0.10 | 0.12 | 0.05 | 0.31 | 30 |

Some of the variances (RECALL and PER-CEIVED_SUCCESS particularly) involved in the simulation are relatively small. This may cause convergence problems for the nlme library. We have followed the recommendations in **https://stat.ethz.ch/pipermail/r-help/2008-June/164797.html** to avoid failures. But we cannot be one hundred per cent sure that false convergence (8) problems will not show up. Even so, the power plots are consistent with those of other variables with similar variances. We are thus confident that the plots are correct.

The Figures B1 to B10 show how the statistical power increases with sample size. The required sample size to achieve 80% power is easy to identify.

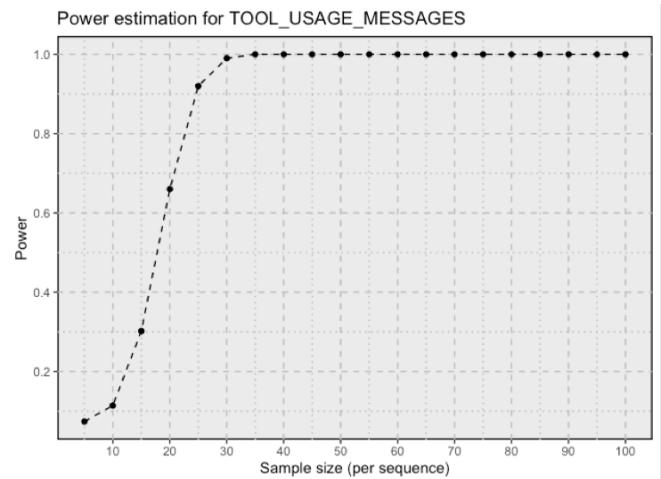

Fig. B3. Power estimation for incremental sample sizes (TOOL_USAGE_MESSAGES response variable).

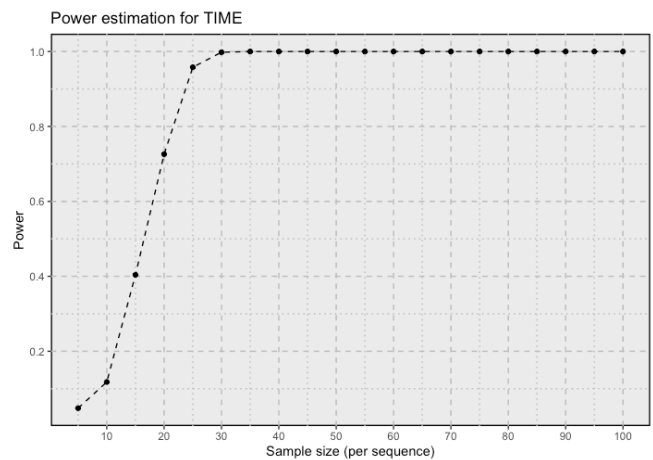

Fig. B1. Power estimation for incremental sample sizes (TIME response variable).

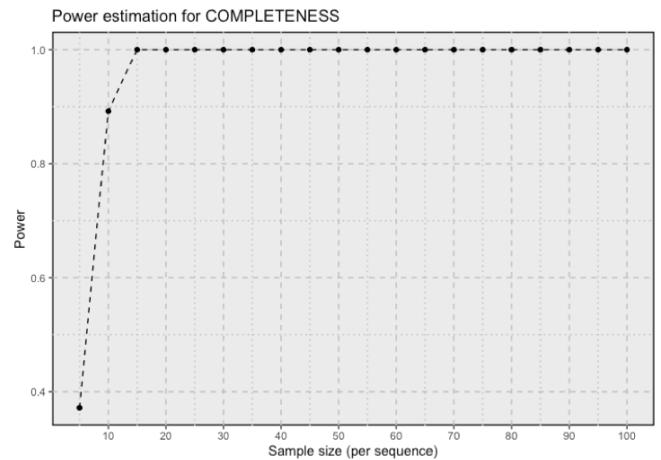

Fig. B4. Power estimation for incremental sample sizes (COMPLETENESS response variable).

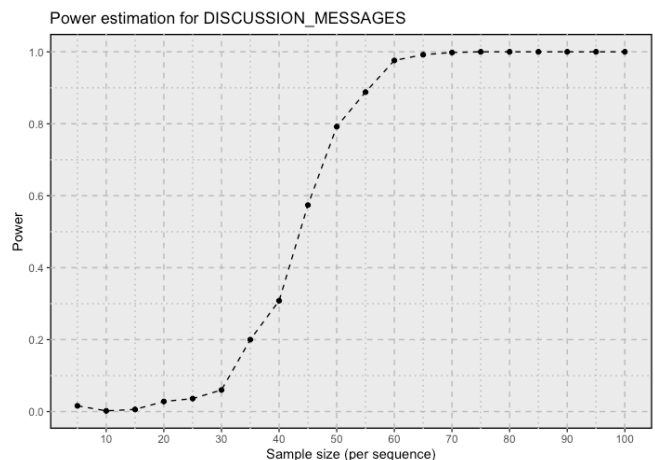

Fig. B2. Power estimation for incremental sample sizes (DISCUSSION_MESSAGES response variable).

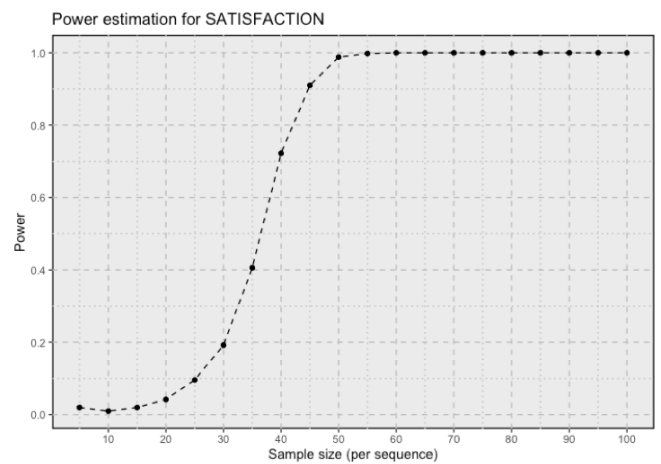

Fig. B5. Power estimation for incremental sample sizes (SATISFACTION response variable).

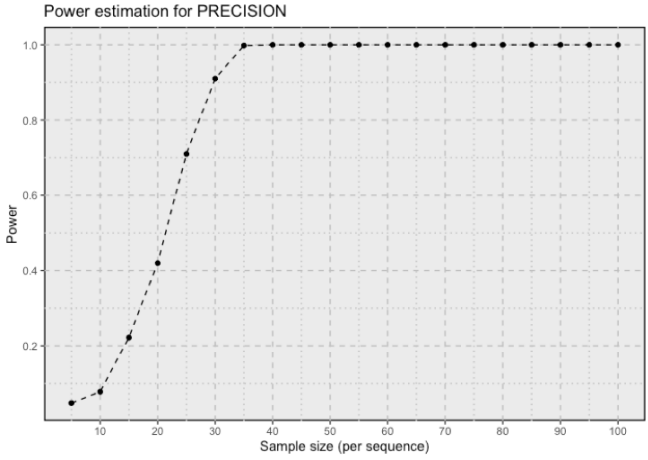

Fig. B6. Power estimation for incremental sample sizes (PRECISION response variable).

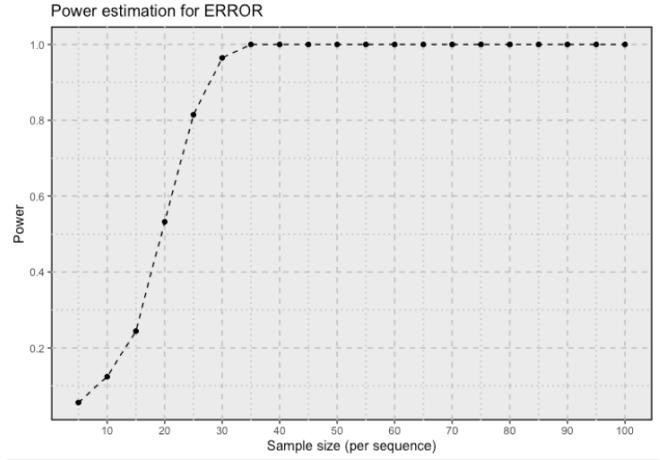

Fig. B9. Power estimation for incremental sample sizes (ERROR response variable).

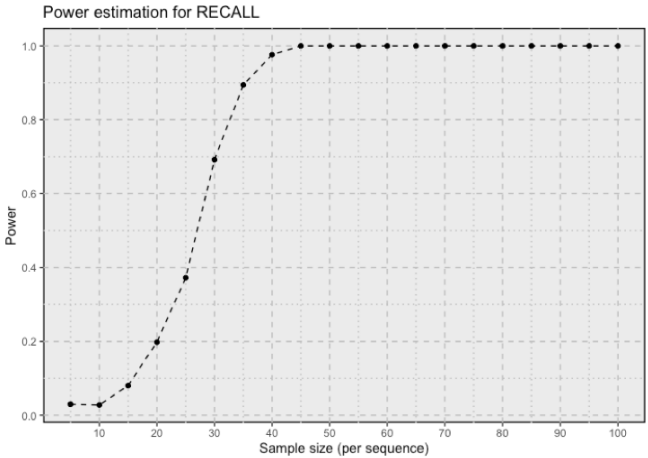

Fig. B7. Power estimation for incremental sample sizes (RECALL response variable).

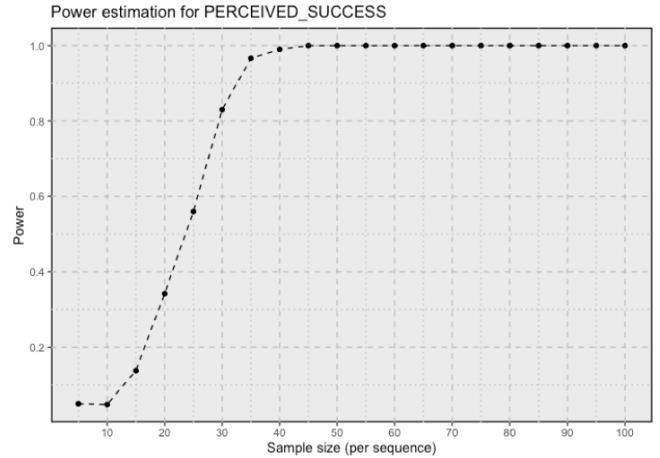

Fig. B10. Power estimation for incremental sample sizes (PERCEIVED_SUCCESS response variable).

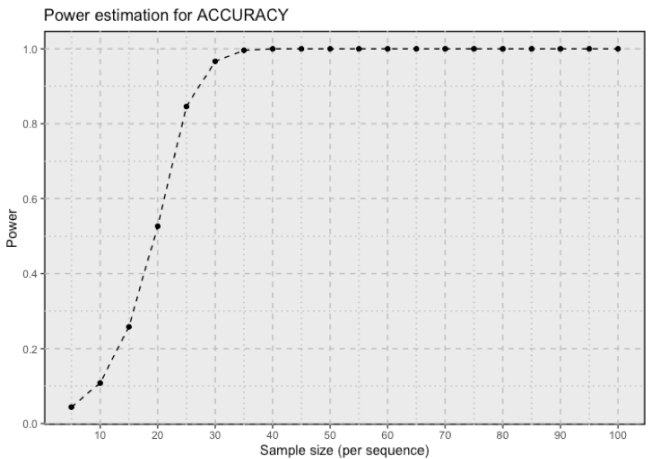

Fig. B8. Power estimation for incremental sample sizes (ACCURACY response variable).

## APPENDIX C    CUMULATIVE META-ANALYSIS

This appendix shows the cumulative meta-analyses that we performed on the treatment estimates for all the metrics within the family. In Figures C1 to C10, we observe that, generally, the 95% confidence intervals get narrower, meaning that the pooled results become more and more precise as the results pile up. There are some exceptions. For instance, the SATISFACTION response variable exhibits wider confidence intervals with three than with two experiments. The confidence intervals could be misleading if the overlap between confidence intervals is interpreted in terms of statistical significance. When the confidence intervals do not overlap, the difference between treatments is statistically significant. However, a small overlap is usually significant as well (see https://stats.stackexchange.com/questions/18215/relation-between-confidence-interval-and-testing-statistical-hypothesis-for-t-te/18259#18259 for further information). To avoid misinterpretations, we have added a label to each analysis indicating whether it is significant. This helps to appreciate the cumulative knowledge discovered by the sequence of experiments.

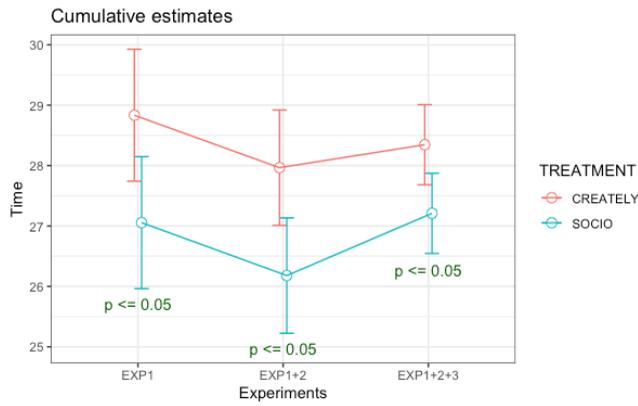

Fig. C1. Cumulative meta-analysis treatment estimates (TIME response variable).

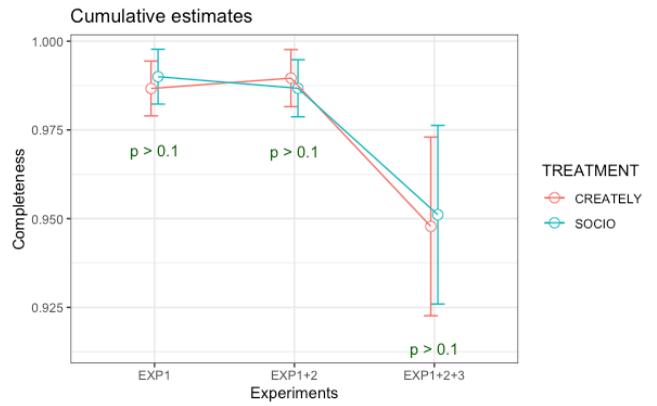

Fig. C4. Cumulative meta-analysis treatment estimates (COMPLETENESS response variable).

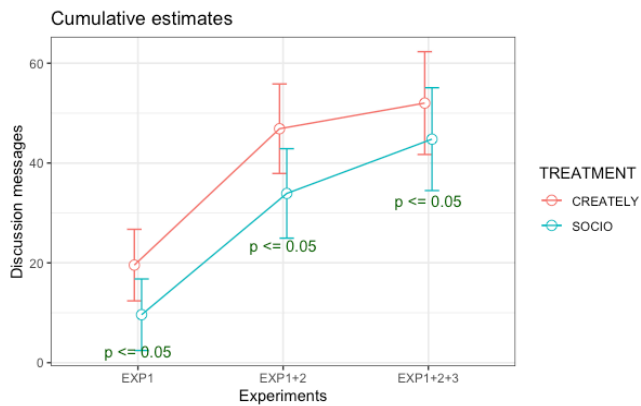

Fig. C2. Cumulative meta-analysis treatment estimates (DISCUSSION_MESSAGES response variable).

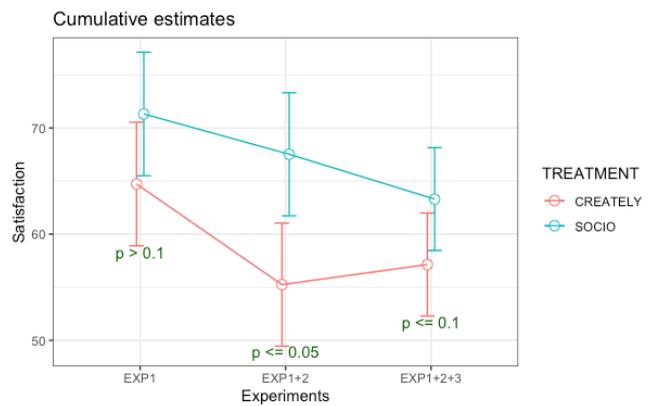

Fig. C5. Cumulative meta-analysis treatment estimates (SATISFACTION response variable).

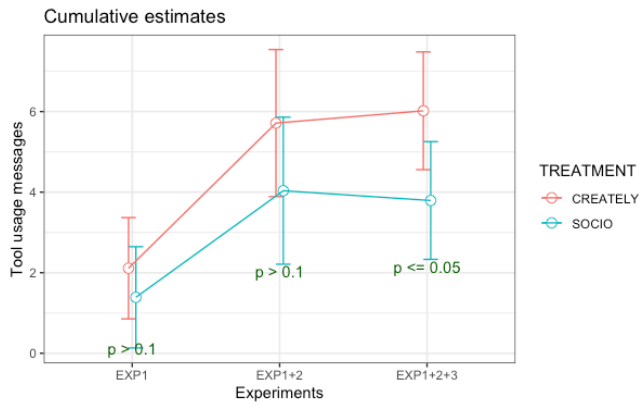

Fig. C3. Cumulative meta-analysis treatment estimates (TOOL_USAGE_MESSAGES response variable).

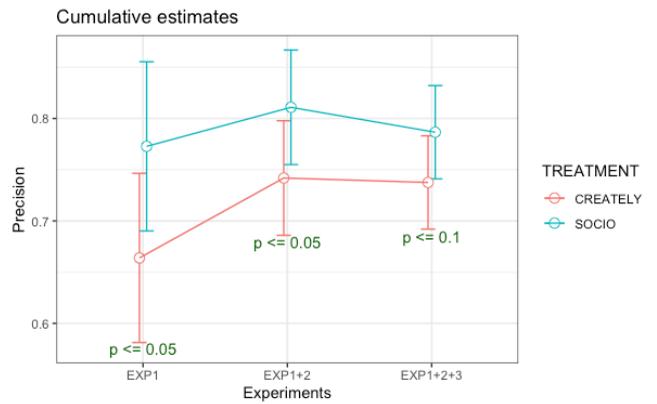

Fig. C6. Cumulative meta-analysis treatment estimates (PRECISION response variable).

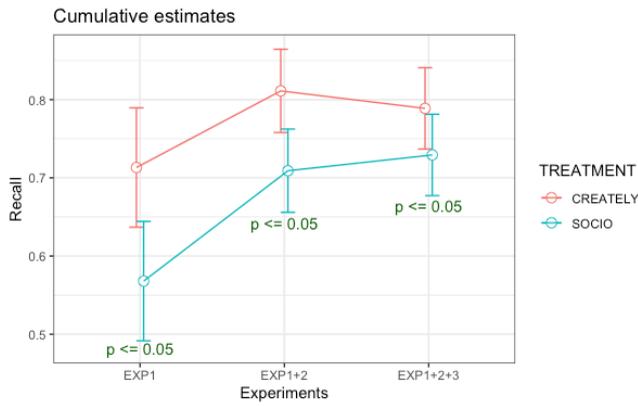

Fig. C7. Cumulative meta-analysis treatment estimates (RECALL response variable).

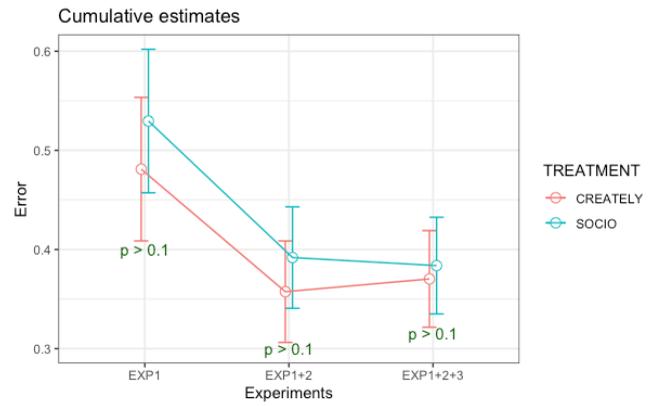

Fig. C9. Cumulative meta-analysis treatment estimates (ERROR response variable).

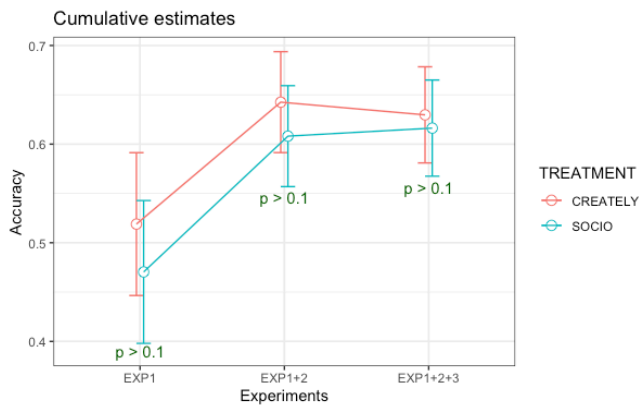

Fig. C8. Cumulative meta-analysis treatment estimates (ACCURACY response variable).

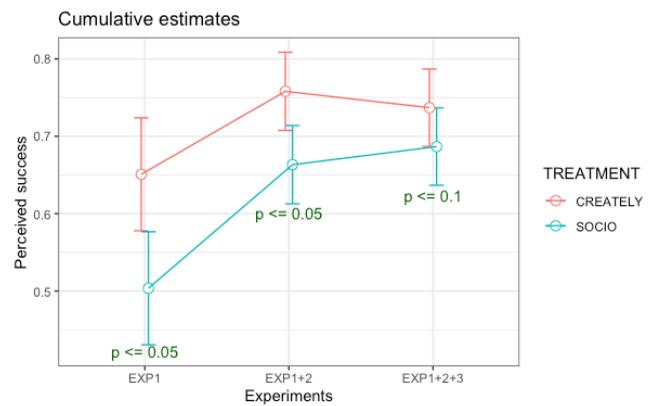

Fig. C10. Cumulative meta-analysis treatment estimates (PERCEIVED_SUCCESS response variable).